\documentclass[preprint]{elsarticle}

\usepackage{lineno,hyperref,amsmath,amsfonts,multirow,color}
\modulolinenumbers[1]
\usepackage[margin=3cm]{geometry}
\usepackage{algorithm}
\usepackage{algpseudocode}
\journal{arXiv}

\begin{document}

\begin{frontmatter}

\title{Exact solutions for diffusive transport on heterogeneous growing domains}

\author{Stuart T. Johnston\corref{correspondingauthor}}
\address{School of Mathematics and Statistics, The University of Melbourne, Australia.}
\cortext[correspondingauthor]{Corresponding author}
\ead{stuart.johnston@unimelb.edu.au}

\author{Matthew J. Simpson}
\address{School of Mathematical Sciences, Queensland University of Technology, Brisbane, Australia.}

\begin{abstract}
From the smallest biological systems to the largest cosmological structures, spatial domains undergo expansion and contraction. Within these growing domains, diffusive transport is a common phenomenon. Mathematical models have been widely employed to investigate diffusive processes on growing domains. However, a standard assumption is that the domain growth is spatially uniform. There are many relevant examples where this is not the case, such as the colonisation of growing gut tissue by neural crest cells. As such, it is not straightforward to disentangle the individual roles of heterogeneous growth and diffusive transport. Here we present exact solutions to models of diffusive transport on domains undergoing spatially non-uniform growth. The exact solutions are obtained via a combination of transformation, convolution and superposition techniques. We verify the accuracy of these solutions via comparison with simulations of a corresponding lattice-based random walk. We explore various domain growth functions, including linear growth, exponential growth and contraction, and oscillatory growth. Provided the domain size remains positive, we find that the derived solutions are valid. The exact solutions reveal the relationship between model parameters, such as the diffusivity and the type and rate of domain growth, and key statistics, such as the survival and splitting probabilities.
\end{abstract}


\end{frontmatter}

\newpage
\section{Introduction}

The expansion and contraction of spatial domains occurs throughout nature \cite{bers2002,landman2007,riess1998,shraiman2005}. At the cosmological scale, the universe is undergoing an accelerating expansion \cite{riess1998}; at the organism scale, regular development requires tissue growth \cite{shraiman2005}; at the cellular scale, cardiomyocytes expand and contract with every heartbeat \cite{bers2002}. Diffusive processes regularly occur within or along such domains. Tissue growth involves the migration of cell populations, which can be considered as a diffusive process \cite{landman2007}. Throughout the intracellular cardiomyocyte environment, chemical species diffuse while driving cellular function \cite{bers2002}. Regulation of diffusive processes within expanding and contracting domains can be critical to healthy function and form. The failure of neural crest cells to colonise growing gut tissue during development leads to Hirschsprung's disease, which is a life-threatening birth defect characterised by a partial or full absence of the enteric nervous system \cite{landman2007}. \\

While a common assumption, domain expansion and contraction is not necessarily a spatially uniform process \cite{binder2008,crampin2002b,yates2012}. Binder \emph{et al.} \cite{binder2008} demonstrate that the growth rate of quail gut tissue during development is different between the two midgut regions and the hindgut region. Accordingly, the migration of quail neural crest cells occurs on a domain that undergoes spatially non-uniform growth. Multilayered drug-loaded nanocapsules exhibit temperature-dependent swelling, which alters the domain through which drug molecules diffuse \cite{zavgorodnya2017}.  \\

Mathematical models are widely used to investigate diffusive processes \cite{crank1979}. However, in the context of diffusive processes on multiple growing domains, previous investigations have typically focused on either a single uniformly growing domain \cite{crampin1999,crampin2002a,landman2003,le2018,klika2017,krapivsky1996,mclean2004,ross2016,ryabov2015,simpson2015a,simpson2015c,yuste2016} or multiple non-growing domains \cite{carr2016,carr2018,carr2019}. For example, Crampin \emph{et al.} \cite{crampin1999,crampin2002a} investigate the conditions sufficient for pattern formation to occur on a uniformly growing domain, which is relevant for biological morphogenesis. Simpson \emph{et al.} \cite{simpson2015b,simpson2015c}, Ryabov \emph{et al.} \cite{ryabov2015} and Yuste \emph{et al.} \cite{yuste2016} each present exact results for specific models of diffusive processes on a uniformly growing domain. The exact results allow key statistics such as the survival probability (the probability that an individual has yet to cross the boundary of an expanding domain) to be readily calculated \cite{redner2001,simpson2015b,simpson2015c,yuste2016}. Approaches have been presented to investigate diffusive processes in spatially heterogeneous non-growing domains \cite{carr2019}, including homogenisation techniques and exact solutions for multilayered piecewise homogeneous domains \cite{carr2018}. One approach for representing spatially non-uniform domain growth is via multiple domains that exhibit piecewise uniform growth \cite{binder2008,crampin2002b}. Numerical results have been presented in the context of pattern formation during non-uniform domain growth, where the non-uniform growth is a piecewise uniform process \cite{crampin2002b}. While both single growing domain problems and multiple non-growing domain problems are both well-studied, there is a lack of exact results detailing the dynamics of diffusive processes on multiple growing domains. Accordingly, it is unclear how spatially non-uniform domain growth may inhibit or enhance diffusive transport. Understanding this interplay between diffusive transport and non-uniform domain growth may provide insight into why normal development processes succeed or fail, and allow the calculation of the effective release rate of therapeutics from nanocapsules. \\

Here we introduce new exact solutions for density profiles, survival probabilities and splitting probabilities for diffusive processes on multiple growing domains. The solutions are obtained via a combination of transformation, convolution and superposition techniques. We demonstrate that the solutions are valid through comparison with density profiles obtained via repeated realisations of a corresponding lattice-based random walk. We show that the solution profiles can exhibit jump discontinuities at inter-domain boundaries and investigate how model parameters, such as the domain growth rate, diffusivity and initial location, influence the survival and splitting probabilities. Critically, each of these parameters can vary between domains, which gives rise to rich behaviour that is not possible on a single uniformly growing domain. We investigate a suite of domain evolution functions, including linear growth \cite{simpson2015c}, exponential growth and decay \cite{barrass2006,binder2009,simpson2015c,yates2014}, and oscillatory evolution \cite{nilsson2020} and show that the exact solutions are valid, provided that the domain size does not reduce to zero.

\begin{figure}
\begin{center}
\includegraphics[width=1.0\textwidth]{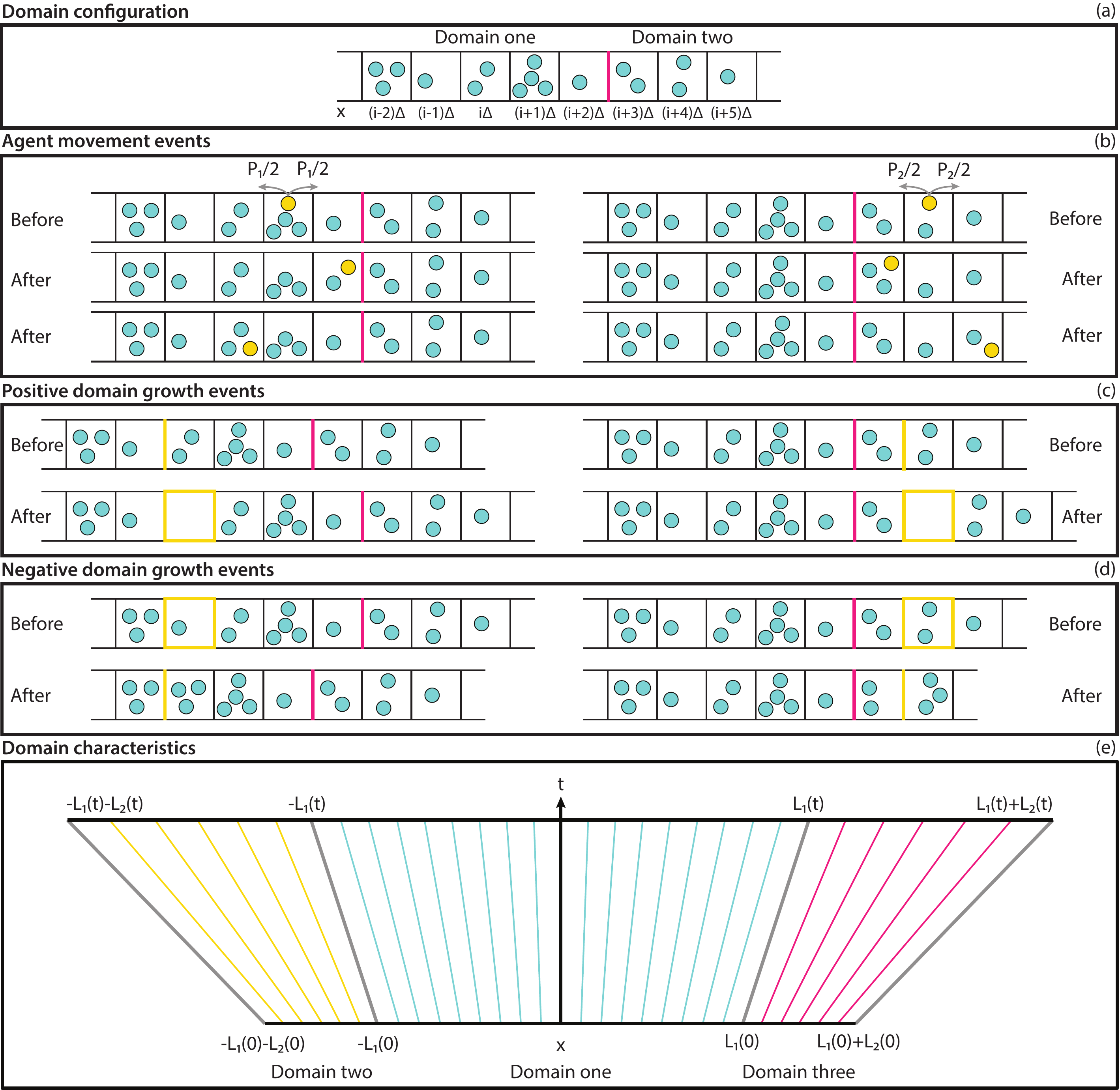}
\caption{(a) Configuration of the domain in the discrete model. Agents exist on a lattice of width $\Delta$. The pink line denotes the boundary between two domains. (b) Lattice configuration before and after possible movement events for agents in either domain. The agent that moves is highlighted in orange. (c) Lattice configuration before and after possible positive domain growth events. The location where the new lattice site is added is highlighted in orange. (d) Lattice configuration before and after possible negative domain growth events. The location where the lattice site is removed is highlighted in orange. (e) Space-time diagram of a growing domain, highlighting the expansion of space. The location of the boundaries are shown in grey. Characteristics for domain one, two and three are shown in cyan, orange and pink, respectively. For visual clarity we present characteristics for a linearly growing domain; in practice, other types of domain growth are possible.}
\label{F1}
\end{center}
\end{figure}
\newpage 
\section{Model}

\subsection{Agent-based discrete model}

We implement a one-dimensional position-jump random walk model on a growing domain \cite{simpson2015c}. Agents in the random walk are located on discrete lattice sites at $x = i\Delta, \ i \in \{-\lfloor L(j\delta t)/\Delta \rfloor, \ \ldots, \ \lfloor L(j\delta t)/\Delta\rfloor\}$ where $L(j\delta t)$ is the half-length of the growing domain after $j \in \mathcal{N}_0$ timesteps of duration $\delta t$ and $\Delta$ is the lattice size. There are no restrictions on the number of agents that can occupy a single lattice site; we note other investigations impose such a restriction \cite{ross2016,ross2017}. Agents located on the site at $x$ randomly move to one of the two nearest-neighbour sites at $x\pm \Delta$ with probability $P/2$ in a timestep. If selected, each agent will therefore move with probability $P$ in a timestep. We select $N(j\delta t)$ agents at random, with replacement, each timestep, where $N(j\delta t)$ is the number of agents on the lattice. If an agent crosses the boundary at $x = \pm \Delta \lfloor L(j\delta t)/\Delta \rfloor $ (i.e. the agent moves to a ``ghost'' lattice site at $x = \pm \Delta \lfloor L(j\delta t)/\Delta + 1\rfloor$) then the agent is determined to have left the domain, and the agent is removed from the random walk at the end of the timestep. \\

We first consider a single domain undergoing spatially-uniform growth. That is, each location in space experiences uniform growth and there is a single growth rate across the domain \cite{simpson2015c}. We implement positive domain growth in the random walk as follows. When $\lfloor L(j\delta t)/\Delta \rfloor > \lfloor L((j-1)\delta t)/\Delta \rfloor$, we randomly select a lattice site $k \in \{1, \ \ldots,\ \lfloor L((j-1)\delta t)/\Delta \rfloor\}$. We insert a new lattice site at $x = k\Delta$, and hence all agents located at sites $i \geq k$ are displaced a distance $\Delta$ in the positive direction \cite{simpson2015c,yates2014}. \textcolor{black}{This implies that the $i$th lattice site translates a distance $\Delta$ with probability $i/\lfloor L((j-1)\delta t)/\Delta \rfloor$ after a domain growth event occurs. Domain growth events occur at a rate proportional to $\text{d}L/\text{d}t$ and hence the product of the probability of translation, the translation distance, and the domain growth event rate can be considered as a discrete velocity field.} We repeat this process for $k \in \{-\lfloor L((j-1)\delta t)/\Delta \rfloor,\ \ldots,\ -1\}$, noting that this results in a displacement of $\Delta$ in the negative direction for agents located at sites $i \leq k$. New lattice sites initially contain zero agents. For negative domain growth (i.e. contraction or shrinkage), when $\lfloor L(j\delta t)/\Delta \rfloor < \lfloor L((j-1)\delta t)/\Delta \rfloor$, we randomly select a lattice site $k \in \{1, \ \ldots,\ \lfloor L((j-1)\delta t)/\Delta \rfloor\}$. We remove the lattice site at $x = k\Delta$, and hence all agents located at sites $i > k$ are displaced a distance $\Delta$ in the negative direction \cite{yates2014}. Agents located at the removed site $k$ are not displaced. This process is repeated for $k \in \{-\lfloor L((j-1)\delta t)/\Delta \rfloor,\ \ldots,\ -1\}$, where there is a displacement of $\Delta$ in the positive direction for agents located at sites $i < k$. A schematic of movement and domain growth events is presented in Figure \ref{F1}. We choose domain growth functions such that $L(j\delta t) = 0$ does not occur. \\

We next consider multiple growing domains, where the growth within each individual domain is spatially uniform. There is a series of adjacent domains (Figure \ref{F1}), each of which grows at a (potentially distinct) uniform rate. Accordingly, the growth rate of the entire domain can be non-uniform. Here agents randomly move in the positive and negative directions with probability $P_g/2$ where $g \in \{1,\ \ldots, \ G\}$ and there are $2G-1$ adjacent, non-overlapping domains in total. Here we always examine domains that are symmetric around $x = 0$ (Figure \ref{F1}(e)). We now have $L(j\delta t) = \sum_{g=1}^{G} L_g(j\delta t)$. New lattice sites are inserted into individual domains at random when $\lfloor L_g(j\delta t)/\Delta \rfloor > \lfloor L_g((j-1)\delta t)/\Delta \rfloor$ and are removed from the individual domains at random when $\lfloor L_g(j\delta t)/\Delta \rfloor < \lfloor L_g((j-1)\delta t)/\Delta \rfloor$ according to the processes described above. \\

We calculate a number of key statistics from individual realisations of the random walk, and perform numerous realisations of the random walk to generate representative average behaviour. Specifically, we calculate the average agent density
\begin{equation*}
\overline{C}(i\Delta,j\delta t) = \frac{1}{\Delta M N} \sum_{m=1}^{M} N_m(i, j), \ i \in \bigg\{-\sum_{g=1}^G\bigg\lfloor \frac{L_g(j\delta t)}{\Delta} \bigg\rfloor, \ \ldots, \ \sum_{g=1}^G\bigg\lfloor \frac{L_g(j\delta t)}{\Delta} \bigg\rfloor\bigg\}, \ j \in \mathcal{N}_0,
\end{equation*}
where $M$ is the number of identically-prepared realisations of the random walk performed, $N$ is the initial number of agents in the random walk, and $N_m(i, j)$ is the number of agents located at site $i$ after $j$ timesteps \textcolor{black}{in the $m$th realisation of the random walk. In practice, in the numerical implementation of the random walk, we track the position of each agent in the random walk, and calculate $N_m(i,j)$ from this information (Supplementary Information)}. We again note that agents are removed from the domain at the end of a timestep if they are located at the ``ghost'' sites at  $x = \pm \Delta(1+\sum_{g=1}^G \lfloor L_g(j\delta t)/\Delta \rfloor )$. We calculate the average survival probability, which represents the proportion of the initial agents that remain on the domain after $j$ timesteps,
\begin{equation*}
\overline{S}(j \delta t) = 1 - \frac{1}{M N} \sum_{m=1}^{M} \sum_i N_m(i, j), \ i \in \bigg\{-\sum_{g=1}^G\bigg\lfloor \frac{L_g(j\delta t)}{\Delta} \bigg\rfloor, \ \ldots, \ \sum_{g=1}^G\bigg\lfloor \frac{L_g(j\delta t)}{\Delta} \bigg\rfloor\bigg\}, \ j \in \mathcal{N}_0.
\end{equation*}
From these time-dependent statistics we can calculate relevant long time statistics such as the average splitting probabilities, which are the proportion of the agents that ever cross the negative and positive boundaries
\begin{equation*}
\overline{S}^-(j\delta t) =  \frac{1}{M N} \sum_{k=1}^j \sum_{m=1}^{M} N_m\bigg(-1-\sum_{g=1}^G\bigg\lfloor \frac{L_g(k\delta t)}{\Delta}\bigg\rfloor, k\bigg),
\end{equation*}
and
\begin{equation*}
\overline{S}^+(j\delta t) = \frac{1}{M N} \sum_{k=1}^j \sum_{m=1}^{M} N_m\bigg(1+\sum_{g=1}^G\bigg\lfloor \frac{L_g(k\delta t)}{\Delta}\bigg\rfloor, k\bigg).
\end{equation*}

Note that here, due to the growing domain, there is no guarantee that all agents will leave the domain as $j \to \infty$. This is in contrast to the non-growing case where all agents eventually leave the domain. The average fraction of agents remaining on the domain at steady state is
\begin{equation*}
\overline{\psi} = \lim_{j \to \infty} \overline{S}(j \delta t) = 1- \overline{\theta}^- - \overline{\theta}^+ = \lim_{j \to \infty} \Big[1 - \overline{S}^-(j \delta t) - \overline{S}^+(j \delta t)\Big]
\end{equation*}
In practice, we must eventually terminate the simulation of the random walk, and hence $j \in \{0, \ \ldots, \ j_{\text{max}}\}$ where $j_{\text{max}}$ is chosen to be sufficiently large that the long-time observations approximate the $t \to \infty$ steady state limit.

\subsection{Continuum model}

Previous studies have derived the relationship between a velocity field $v(x,t)$ that translates each point $0 \leq x \leq L(t)$ in a domain, and the overall growth rate of the domain due to the expansion of each infinitesimal width of space \cite{landman2003},
\begin{equation*}
\frac{\text{d}L(t)}{\text{d}t} = \int_0^{L(t)} \frac{\partial v(x,t)}{\partial x} \ \text{d} x.
\end{equation*}
We note that in all cases here we consider a symmetric growing domain centred around $x = 0$, that is, $v(x,t) = -v(-x,t)$. To account for the multiple growing domains we choose a velocity field that is piecewise defined for the separate domains. Accordingly,
\begin{equation*}
\frac{\text{d}L(t)}{\text{d}t} = \sum_{g=1}^G \frac{\text{d}L_g(t)}{\text{d}t} = \int_0^{L_1(t)} \frac{\partial v(x,t)}{\partial x} \ \text{d} x + \sum_{g=1}^{G-1} \int_{B_{g}(t)}^{B_{g+1}(t)} \frac{\partial v(x,t)}{\partial x} \ \text{d} x,
\end{equation*}
where, for ease of notation, we define the position of the $g$th moving boundary
\begin{equation}
B_g(t) = \sum_{i=1}^{g} L_i(t), \ g \in\{1,\ \ldots,\ G\}, \label{Eq:Boundaries}
\end{equation}
with $B_0(t) = 0$. Here we select spatially-uniform growth within each individual domain. That is, $v(x,t)$ is piecewise and chosen such that $\partial v/\partial x$ is independent of $x$ within an individual domain, though we note there may be a temporal dependence,
\begin{equation*}
\frac{\partial v(x,t)}{\partial x} = \sigma_g(t) = \frac{1}{L_g(t)}\frac{\text{d}L_g(t)}{\text{d}t}, \ B_{g-1}(t) < x < B_g(t), \ g \in\{1,\ \ldots,\ G\}.
\end{equation*}
The velocity field is therefore
\begin{equation}
v(x,t) = 
 \frac{1}{L_g(t)}\Big(x-B_{g-1}(t)\Big)\frac{\text{d}L_g(t)}{\text{d}t} + \sum_{i=1}^{g-1}\frac{\text{d}L_i(t)}{\text{d}t}, \ \ B_{g-1}(t) < x < B_g(t), \ g \in\{1, \ \ldots, \ G\}, \label{Eq:GeneralVelocity}
\end{equation}
and recalling that $v(x,t) = -v(-x,t)$ with $v(0,t) = 0$. We observe that the velocity field at the boundary between domains is the sum of the domain growth rates for all domains to that point (Figure \ref{F1}(e)). \\

We now consider the evolution of a mass that undergoes linear diffusion on this growing domain. The governing equation for the mass density $C(x,t)$ is \cite{simpson2015c}
\begin{equation}
\frac{\partial C(x,t)}{\partial t} = \frac{\partial}{\partial x}\bigg(D(x)\frac{\partial C(x,t)}{\partial x}\bigg) - \frac{\partial}{\partial x}\bigg(v(x,t)C(x,t)\bigg), \ -L(t) < x < L(t), \label{Eq:GeneralGoverning}
\end{equation}
where the diffusivity $D(x)$ is piecewise defined 
\begin{equation}
D(x) = D_g = 
\lim_{\Delta, \delta t \to 0} \frac{\Delta^2 P_g}{2\delta t}, \ \ B_{g-1}(t) < x < B_{g}(t) \cup -B_{g}(t) < x < -B_{g-1}(t), \ g \in\{1,\ \ldots,\ G\}, \label{Eq:GeneralDiffusivity}
\end{equation} on each individual domain in line with the agent-based model, and $v(x,t)$ is defined \textcolor{black}{in Equation \eqref{Eq:GeneralVelocity}}. To be consistent with the agent-based model we impose 
\begin{equation}
C(-L(t),t) = C(L(t),t) = 0, \label{Eq:OuterBCs}
\end{equation}
that is, any mass that reaches the external boundary is removed from the system. Other boundary conditions could be considered, such as a no-flux boundary condition at one domain boundary as in \cite{simpson2015c}, but such extensions are left for future work. For the boundaries between internal domains, the flux across the boundaries must be conserved and hence
\begin{equation}
- D_g \frac{\partial C}{\partial x}\bigg|_{x=B_g(t)^-} = - D_{g+1} \frac{\partial C}{\partial x}\bigg|_{x=B_g(t)^+}. \label{Eq:InnerBCs}
\end{equation}
Equations \eqref{Eq:Boundaries}-\eqref{Eq:InnerBCs} therefore define the general form of the model for a yet-to-be-specified choice of $L(t)$. Here all functional forms for $L(t)$ are chosen to be smooth functions where $L_g(t) = 0$ does not occur, thereby avoiding finite-time blow-up in the solutions.

\subsubsection{Fixed domain solutions}
To aid in constructing exact solutions to the model for multiple growing domains, we first focus on the simplest case where there is a single domain of fixed size. Here the governing equation reduces to the classical diffusion equation on a fixed domain
\begin{equation*}
\frac{\partial C(x,t)}{\partial t} = D \frac{\partial^2 C(x,t)}{\partial x^2}, \ -L < x < L, \ C(-L,t) = C(L,t) = 0.
\end{equation*}
For a Dirac delta initial condition located at $x = x_0$, the solution obtained via method of images \cite{crank1979} is
\begin{align}
C(x,t) = &\frac{1}{\sqrt{4\pi Dt}}\exp\bigg(-\frac{(x-x_0)^2}{4Dt}\bigg) \nonumber \\ &+ \frac{1}{\sqrt{4\pi Dt}} \Bigg[\sum_{j = 1}^{\infty}(-1)^j  \exp\bigg(-\frac{(x-(-2jL+(-1)^jx_0))^2}{4Dt}\bigg) \nonumber \\ &+ (-1)^j \exp\bigg(-\frac{(x-(2jL+(-1)^jx_0))^2}{4Dt}\bigg)\Bigg].
\label{Eq:FixedDomainSolution}
\end{align}
The fluxes at the two boundaries, which represent the instantaneous rate of mass leaving the domain, are
\begin{align}
F^-(t) = -D \frac{\partial C(x,t)}{\partial x}\bigg|_{x=-L} &= \sum_{j=0}^{\infty}(-1)^j \frac{(2j+1)L+(-1)^jx_0}{\sqrt{4\pi Dt^3}}\exp\bigg(-\frac{((2j+1)L+(-1)^jx_0)^2}{4Dt}\bigg), \label{Eq:SingleFixedLeftFlux} \\
F^+(t) = -D \frac{\partial C(x,t)}{\partial x}\bigg|_{x=L} &= \sum_{j=0}^{\infty} (-1)^j \frac{(2j+1)L-(-1)^jx_0}{\sqrt{4\pi Dt^3}}\exp\bigg(-\frac{((2j+1)L-(-1)^jx_0)^2}{4Dt}\bigg). \label{Eq:SingleFixedRightFlux}
\end{align}
Note that while the solution is an infinite summation, in practice, we can approximate this solution via a truncated summation. In all cases we choose a summation with 10 terms and verify that the solution is not sensitive to the inclusion of additional terms.

\subsubsection{Single growing domain solutions}

We next focus on exact solutions for the case with a single growing domain. Here the governing equation is \cite{simpson2015c}
\begin{equation*}
\frac{\partial C(x,t)}{\partial t} = D\frac{\partial^2 C(x,t)}{\partial x^2} - \frac{\partial}{\partial x}\bigg(v(x,t)C(x,t)\bigg), \ -L(t) < x < L(t), \ C(L(t),t) = C(-L(t),t) = 0,
\end{equation*}
where \textcolor{black}{
\begin{equation*}
v(x,t) = \frac{x}{L(t)}\frac{\text{d}L}{\text{d}t},
\end{equation*}
and $L(t)$ is yet to be specified.} If we can transform this equation to a linear diffusion equation on a fixed domain, we can exploit the various properties of solutions to the linear diffusion equation.  We consider transformations of $C(x,t), \ x$, and $t$ to $C(\xi,T), \ \xi$, and $T$, respectively.  A natural spatial scale is the ratio of the domain size to the spatial variable and hence we introduce
\begin{equation*}
\xi = \frac{x}{L(t)}L(0), \ -L(0) < \xi < L(0).
\end{equation*}
\textcolor{black}{Applying this transformation,} we obtain, after cancellation of the first order terms in $\xi$,
\begin{equation*}
\frac{\partial C(\xi,t)}{\partial t} = \frac{DL(0)^2}{L(t)^2}\frac{\partial^2C(\xi,t)}{\partial \xi^2} - \frac{C(\xi,t)}{L(t)}\frac{\text{d}L(t)}{\text{d}t}.
\end{equation*}
The new source term acts to dilute (concentrate) the mass density in the transformed coordinate system, reflecting the stretched (contracted) nature of the domain for $\text{d}L(t)/\text{d}t > 0$ ($\text{d}L(t)/\text{d}t < 0$). \textcolor{black}{Following previous approaches \cite{crank1979,ranz1979}, we now introduce} a temporal scaling such that the coefficient of the second $\xi$ derivative term is a constant,
\begin{equation*}
T(t) = \int_0^{t} \bigg(\frac{L(0)}{L(s)}\bigg)^2 \ \text{d} s,
\end{equation*}
and hence
\begin{equation*}
\frac{\partial C(\xi,T)}{\partial T} = D\frac{\partial^2C(\xi,T)}{\partial \xi^2} - f(T)C(\xi,T),
\end{equation*}
where 
\begin{equation*}
f(T) = \frac{L(t)}{L(0)^2}\frac{\text{d}L(t)}{\text{d}t}.
\end{equation*}
We now define
\begin{equation*}
C(\xi,T) = U(\xi,T)\exp\bigg(-\int_0^T f(s) \ \text{d}s\bigg).
\end{equation*}
Via a change of variable in the integral we obtain
\begin{equation*}
C(\xi,T) = U(\xi,T)\frac{L(0)}{L(t)},
\end{equation*}
and hence $U(\xi,T)$ satisfies
\begin{equation*}
\frac{\partial U(\xi,T)}{\partial T} = D\frac{\partial^2U(\xi,T)}{\partial \xi^2},  \ -L(0) < \xi < L(0),  \ T > 0,
\end{equation*}
which is the linear diffusion equation on a finite domain. This result is equivalent to the result derived in Simpson \emph{et al.} \cite{simpson2015c}.  As such,  there are various well-known methods of solution for $U(\xi,T)$ and it remains to transform back from the fixed domain solution $U(\xi,T)$ to the growing domain solution $C(x,t)$.  Consider the fundamental solution to the diffusion equation on an infinite domain with a Dirac delta initial condition at $x = x_0$,  which corresponds to $\xi = x_0L(0)/L(t)$,
\begin{equation*}
U(\xi,T) = \frac{1}{\sqrt{4\pi DT}}\exp\bigg(-\frac{\Big(\xi-x_0\frac{L(0)}{L(t)}\Big)^2}{4DT}\bigg), \ -\infty < \xi < \infty,
\end{equation*}
and hence 
\begin{equation*}
C(\xi,T) = \frac{1}{\sqrt{4\pi DT}}\exp\bigg(-\frac{\Big(\xi-x_0\frac{L(0)}{L(t)}\Big)^2}{4DT}\bigg)\frac{L(0)}{L(t)}.
\end{equation*}
Transforming back into the original space and time coordinates,  we have 
\begin{equation*}
C(x,t) = \frac{1}{\sqrt{4\pi DT(t)}}\exp\bigg(-\frac{\Big(x\frac{L(0)}{L(t)}-x_0\Big)^2}{4DT(t)}\bigg)\frac{L(0)}{L(t)}.
\end{equation*}

To satisfy the homogeneous Dirichlet boundary conditions at $x = \pm L(t)$, additional terms are required in the solution to account for the flux through the boundaries and hence
\begin{align}
C(x,t) = &\frac{1}{\sqrt{4\pi DT(t)}}\frac{L(0)}{L(t)}\Bigg[\exp\bigg(-\frac{\Big(x\frac{L(0}{L(t)}-x_0\Big)^2}{4DT(t)}\bigg) \nonumber \\ &+  \Bigg[\sum_{j = 1}^{\infty}(-1)^j  \exp\bigg(-\frac{\Big(x\frac{L(0)}{L(t)}-(-2jL(0)+(-1)^jx_0)\Big)^2}{4DT(t)}\bigg) \nonumber \\ &+ (-1)^j \exp\bigg(-\frac{\Big(x\frac{L(0)}{L(t)}-(2jL(0)+(-1)^jx_0)\Big)^2}{4DT(t)}\bigg)\Bigg].
\label{Eq:GrowingDomainSolution}
\end{align}
\textcolor{black}{We note that other solutions to the diffusion equation could be chosen. This is particularly relevant for initial conditions that are not a Dirac delta function, such as if $C(x,0)$ is chosen to be a Heaviside function, or the difference between two Heaviside functions \cite{crank1979,simpson2015c}. However, as we detail below, the fundamental solution proves insightful for the case with multiple domains.}

\subsubsection{Multiple fixed domain solutions}

\begin{figure}
\begin{center}
\includegraphics[width=1.0\textwidth]{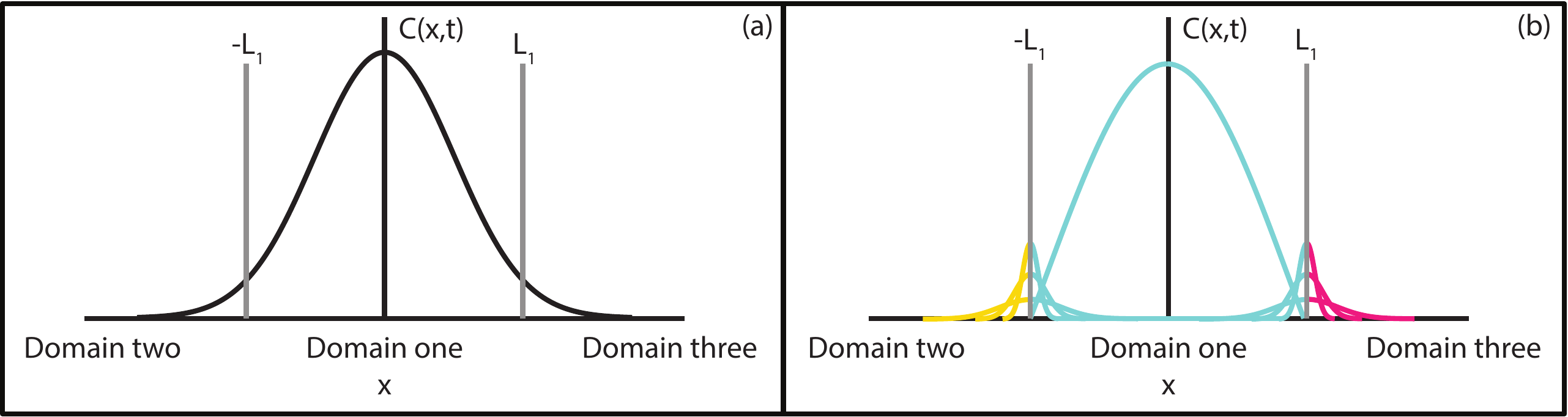}
\end{center}
\caption{(a) Illustration of a solution to the diffusion equation on multiple fixed domains and (b) its constituent components. The multiple solution profiles centred at the boundaries are an illustrative representation of the convolution terms in Equations \eqref{Eq:MultipleFixedDomainsSolution1}, \eqref{Eq:MultipleFixedDomainsSolution3} and \eqref{Eq:MultipleFixedDomainsSolution4}.}
\label{F2}
\end{figure}

Before considering the solutions for the model of multiple growing domains, we now consider exact solutions for the simplified case of multiple fixed domains (Figure \ref{F2}). This allows us to demonstrate the methods for accounting for the flux across internal boundaries, which will be employed to solve the equations for the multiple growing domains case. Here the governing equations are
\begin{align*}
\frac{\partial C_1(x,t)}{\partial t} &= D_1\frac{\partial}{\partial x}\bigg(\frac{\partial C_1(x,t)}{\partial x}\bigg), \ -L_1 < x < L_1, \\
\frac{\partial C_2(x,t)}{\partial t} &= D_2\frac{\partial}{\partial x}\bigg(\frac{\partial C_2(x,t)}{\partial x}\bigg), \ -L_2 - L_1 < x < -L_1, \\
\frac{\partial C_3(x,t)}{\partial t} &= D_2\frac{\partial}{\partial x}\bigg(\frac{\partial C_3(x,t)}{\partial x}\bigg), \ L_1 < x < L_1+L_2,
\end{align*}
where the rate of diffusion in domain two is always the same as in domain three. The flux across the internal boundaries at $x = \pm L_1$ is conserved and hence
\begin{equation*}
-D_1 \frac{\partial C_1}{\partial x}\bigg|_{x=-L_1} = -D_2 \frac{\partial C_2}{\partial x}\bigg|_{x=-L_1}, \qquad -D_1 \frac{\partial C_1}{\partial x}\bigg|_{x=L_1} = -D_2 \frac{\partial C_3}{\partial x}\bigg|_{x=L_1}.
\end{equation*}
Individuals are absorbed at the far boundaries, such that $C_2(-L_2-L_1,t) = 0$ and $C_3(L_1+L_2,t) = 0$. We construct solutions by first solving for $C_1(x,t)$ subject to $C_1(\pm L_1,t) = 0$. We will denote this solution $G(x,t)$, which is the solution to a related problem (with different boundary conditions) that provides the insight necessary to solve the problem for the original boundary conditions. We calculate the flux that leaves the inner domain according to $G(x,t)$, and then re-introduce a fraction of that flux at $x = \pm L_1$. This fraction represents the agents that cross the boundary but ultimately return to the original domain. Due to the linearity property of the diffusion equation, we are able to superimpose solutions that satisfy the governing equations. We have defined the boundary fluxes $F^-(t)$ and $F^+(t)$ in Equations \eqref{Eq:SingleFixedLeftFlux}-\eqref{Eq:SingleFixedRightFlux}. The relevant fraction to be added at the boundary can be determined from conservation of flux. Observing that the solution to $C_1(x,t)$ is composed of the related solution $G(x,t)$ (which, by definition, is zero at $x = \pm L_1$) and the re-introduced sources, the solution at the boundary at $x = L_1$ is
\begin{equation*}
C_1(L_1,t) = \lim_{s\to t} \int_0^s\frac{w_1^+}{\sqrt{\pi D_1(s-\tau)}} F^+(\tau)  \ \text{d}\tau,
\end{equation*}
to first order; see below for a discussion on the timescales over which this solution is accurate. Here $w_1^+$ is both the fraction of the flux that remains in domain one, and a weight that relates the flux at $x = L_1$ in the positive $x$ direction for $G(x,t)$ to the flux for $C_1(x,t)$, that is,
\begin{equation*}
-D_1(1-w_1^+)\frac{\partial G(x,t)}{\partial x}\bigg|_{x=L_1} = -D_1\frac{\partial C_1(x,t)}{\partial x}\bigg|_{x=L_1}.
\end{equation*}
The solution in domain three at the same boundary, which is only composed to the introduced sources, is (to first order)
\begin{equation*}
C_3(L_1,t) = \lim_{s\to t} \int_0^s \frac{w_3^-}{\sqrt{\pi D_2(s-\tau)}} F^+(\tau) \ \text{d}\tau,
\end{equation*}
where $w_3^-$ is the fraction of flux that enters domain three, subject to the restrction that $w_1^+ + w_3^- = 1$. For conservation of flux to hold, we require
\begin{equation*}
w_1^+ = \frac{\sqrt{D_2}}{\sqrt{D_1}+\sqrt{D_2}}, \qquad w_3^- =  \frac{\sqrt{D_1}}{\sqrt{D_1}+\sqrt{D_2}}
\end{equation*}
at $x = L_1$. Similarly, for the boundary at $x = -L_1$, we require 
\begin{equation*}
w_1^- = \frac{\sqrt{D_2}}{\sqrt{D_1}+\sqrt{D_2}}, \qquad w_2^+ =  \frac{\sqrt{D_1}}{\sqrt{D_1}+\sqrt{D_2}}.
\end{equation*}
If $D_1 = D_2$ then it is equally likely for an agent that crosses the boundary to be located in either domain. If $D_1 > D_2$ then agents reside in the outer domains for longer than in the inner domain, and the agent density will be higher in the outer domains; equally, if $D_1 < D_2$ then agents reside in the inner domain for longer, and the agent density will be higher in the inner domain. For a Dirac delta initial condition located at $x  = x_0$ the solution on $-L_1 < x < L_1$ is therefore (Figure \ref{F2})
\begin{align}
C_1(x,t) = G(x,t) + \int_0^t w_1^-F^-(\tau)P_1^-(x,-L_1,t-\tau) + w_1^+F^+(\tau)P_1^+(x,L_1,t-\tau) \ \text{d}\tau,
\label{Eq:MultipleFixedDomainsSolution1}
\end{align}
where 
\begin{align}
G(x,t) = &\frac{1}{\sqrt{4\pi D_1t}}\exp\bigg(-\frac{(x-x_0)^2}{4D_1t}\bigg) \nonumber \\ &+ \frac{1}{\sqrt{4\pi D_1t}} \Bigg[\sum_{j = 1}^{\infty}(-1)^j  \exp\bigg(-\frac{(x-(-2jL_1+(-1)^jx_0))^2}{4D_1t}\bigg) \nonumber \\ &+ (-1)^j \exp\bigg(-\frac{(x-(2jL_1+(-1)^jx_0))^2}{4D_1t}\bigg)\Bigg],
\label{Eq:MultipleFixedDomainsSolution2}
\end{align}
is the solution to the diffusion equation with homogeneous Dirichlet boundary conditions, $P_i^k(x,x_0,t), \ k \in \{-,+\}$ is the fundamental solution to the diffusion equation on domain $i$ with a Dirac delta initial condition at $x = x_0$, a homogeneous Neumann boundary condition at the leftmost boundary, and a homogeneous Dirichlet boundary condition at the rightmost boundary (for $k = -$; the boundary conditions are reversed for $k = +$), and
\begin{equation*}
F^-(t) = -D_1\frac{\partial G}{\partial x}\bigg|_{x = -L_1}, \qquad F^+(t) = -D_1\frac{\partial G}{\partial x}\bigg|_{x=L_1},
\end{equation*}
as before. For example,
\begin{align*}
P_1^-(x,x_0,t) =  &\frac{1}{\sqrt{4\pi D_1 t}}\Bigg[\exp\bigg(-\frac{(x-x_0)^2}{4D_1t}\bigg) \\
& + \sum_{j=1}^{\infty} (-1)^{\lfloor (j+1)/2 \rfloor} \exp\bigg(-\frac{(x-(-2jL_1+(-1)^jx_0)^2}{4D_1t}\bigg) \\
& + \sum_{j=1}^{\infty} (-1)^{\lfloor j/2 \rfloor} \exp\bigg(-\frac{(x-(2jL_1+(-1)^jx_0)^2}{4D_1t}\bigg)\Bigg].
\end{align*}
Note that the mass contained in the solution components corresponding to the reintroduced flux will eventually reach the opposing boundary (i.e. mass reintroduced at $x = L_1$ will reach $x = -L_1$). Due to the homogeneous Dirichlet boundary imposed at $x = -L_1$ for the solution component $P_1^+(x,L_1,t-\tau)$, this implies that there will be an additional flux through the boundary at $x = -L_1$. This could be accounted for in a similar way to the original boundary flux, where the appropriate fraction of the flux is reintroduced on either side of the boundary via a convolution term. The solution would therefore involve a double convolution component. However, here we consider problems where the average timescale for an individual to cross both internal boundaries is beyond the timescale of interest, so we neglect this additional term. The solutions for $-L_1 -L_2 < x < -L_1$ and $L_1 < x < L_1+L_2$ are
\begin{align}
C_2(x,t) &= \int_0^t w_2^+F^-(\tau)P_2^+(x,-L_1,t-\tau) \ \text{d}\tau, \label{Eq:MultipleFixedDomainsSolution3} \\ 
C_3(x,t) &= \int_0^t w_3^-F^+(\tau)P_3^-(x,L_1,t-\tau) \ \text{d}\tau. \label{Eq:MultipleFixedDomainsSolution4}
\end{align}
The solution at the boundary exhibits the jump condition \cite{gudnason2018}
\begin{equation*}
\frac{C_1(L,t)}{C_3(L,t)} = \frac{D_2}{D_1}.
\end{equation*}

\subsubsection{Multiple growing domain solutions}

We next focus on obtaining exact solutions for the case with multiple growing domains. For $G = 2$, the governing equations are
\begin{align*}
\frac{\partial C_1(x,t)}{\partial t} &= D_1\frac{\partial}{\partial x}\bigg(\frac{\partial C_1(x,t)}{\partial x}\bigg) - \frac{\partial}{\partial x}\bigg(v_1(x,t)C_1(x,t)\bigg), \ -L_1(t) < x < L_1(t), \\
\frac{\partial C_2(x,t)}{\partial t} &= D_2\frac{\partial}{\partial x}\bigg(\frac{\partial C_2(x,t)}{\partial x}\bigg) - \frac{\partial}{\partial x}\bigg(v_2(x,t)C_2(x,t)\bigg), \ -L_2(t) - L_1(t) < x < -L_1(t), \\
\frac{\partial C_3(x,t)}{\partial t} &= D_2\frac{\partial}{\partial x}\bigg(\frac{\partial C_3(x,t)}{\partial x}\bigg) - \frac{\partial}{\partial x}\bigg(v_3(x,t)C_3(x,t)\bigg), \ L_1(t) < x < L_1(t)+L_2(t)
\end{align*}
where $L_1(t)$ and $L_2(t)$ are yet to be specified and $v_3(x,t) = -v_2(-x,t)$ such that the domain is symmetric. We consider transformations of $C_1(x,t)$,  $C_2(x,t)$,  $C_3(x,t)$, $x$, and $t$ to $C_1(\xi_1,T_1)$,  $C_2(\xi_2,T_2)$, $C_3(\xi_3,T_2)$, $\xi_1$,  $\xi_2$,  $\xi_3$, $T_1$, and $T_2$.  The three spatial transformations are
\begin{equation*}
\xi_1 = \bigg(\frac{x}{L_1(t)}\bigg)L_1(0), \qquad \xi_2 = \bigg(\frac{x-L_1(t)}{L_2(t)}\bigg)L_2(0) + L_1(0), \qquad \xi_3 = \bigg(\frac{x+L_1(t)}{L_2(t)}\bigg)L_2(0) - L_1(0).
\end{equation*}
Here we have 
\begin{align*}
v_1(x,t) = \bigg(\frac{x}{L_1(t)}\bigg)\frac{\text{d}L_1}{\text{d}t}, &\qquad v_2(x,t) = \bigg(\frac{x-L_1(t)}{L_2(t)}\bigg)\frac{\text{d}L_2}{\text{d}t} + \frac{\text{d}L_1}{\text{d}t},  \\ \qquad v_3(x,t) &= \bigg(\frac{x+L_1(t)}{L_2(t)}\bigg)\frac{\text{d}L_2}{\text{d}t} - \frac{\text{d}L_1}{\text{d}t}.
\end{align*}
Applying the transformations, we obtain
\begin{align*}
\frac{\partial C_1(\xi_1,t)}{\partial t} &= D_1\bigg(\frac{L_1(0)}{L_1(t)}\bigg)^2 \frac{\partial^2C_1(\xi_1,t)}{\partial \xi_1^2} - \frac{1}{L_1(t)}\frac{\text{d}L_1}{\text{d}t}C_1(\xi_1,t), \ -L_1(0) < \xi_1 < L_1(0), \\
\frac{\partial C_2(\xi_2,t)}{\partial t} &= D_2\bigg(\frac{L_2(0)}{L_2(t)}\bigg)^2 \frac{\partial^2C_2(\xi_2,t)}{\partial \xi_2^2} - \frac{1}{L_2(t)}\frac{\text{d}L_2}{\text{d}t}C_2(\xi_2,t), \ -L_2(0)-L_1(0) < \xi_2 < -L_1(0), \\
\frac{\partial C_3(\xi_3,t)}{\partial t} &= D_2\bigg(\frac{L_2(0)}{L_2(t)}\bigg)^2 \frac{\partial^2C_3(\xi_3,t)}{\partial \xi_3^2} - \frac{1}{L_2(t)}\frac{\text{d}L_2}{\text{d}t}C_3(\xi_3,t), \ L_1(0) < \xi_3 < L_1(0)+L_2(0).
\end{align*}
The time transformations are
\begin{equation*}
T_1(t) = \int_0^t \bigg(\frac{L_1(0)}{L_1(s)}\bigg)^2 \ \text{d}s, \qquad T_2(t) = \int_0^t \bigg(\frac{L_2(0)}{L_2(s)}\bigg)^2 \ \text{d}s,
\end{equation*}
which give
\begin{align*}
\frac{\partial C_1(\xi_1,T_1)}{\partial T_1} &= D_1\frac{\partial^2C_1(\xi_1,T_1)}{\partial \xi_1^2} - \frac{L_1(t)}{L_1(0)^2}\frac{\text{d}L_1}{\text{d}t}C_1(\xi_1,T_1), \ -L_1(0) < \xi_1 < L_1(0), \\
\frac{\partial C_2(\xi_2,T_2)}{\partial T_2} &= D_2 \frac{\partial^2C_2(\xi_2,T_2)}{\partial \xi_2^2} - \frac{L_2(t)}{L_2(0)^2}\frac{\text{d}L_2}{\text{d}t}C_2(\xi_2,T_2), \ -L_2(0)-L_1(0) < \xi_2 < -L_1(0), \\
\frac{\partial C_3(\xi_3,T_2)}{\partial T_2} &= D_2 \frac{\partial^2C_3(\xi_3,T_2)}{\partial \xi_3^2} - \frac{L_2(t)}{L_2(0)^2}\frac{\text{d}L_2}{\text{d}t}C_3(\xi_3,T_2), \ L_1(0) < \xi_3 < L_1(0)+L_2(0). 
\end{align*}
If we define
\begin{equation*}
f_1(T_1) = \frac{L_1(t)}{L_1(0)^2}\frac{\text{d}L_1}{\text{d}t}, \qquad f_2(T_2) = \frac{L_2(t)}{L_2(0)^2}\frac{\text{d}L_2}{\text{d}t},
\end{equation*}
and
\begin{equation*}
C_1(\xi_1,T_1) = U_1(\xi_1,T_1)\frac{L_1(0)}{L_1(t)}, \qquad C_2(\xi_2,T_2) = U_2(\xi_2,T_2)\frac{L_2(0)}{L_2(t)}, \qquad C_3(\xi_3,T_2) = U_3(\xi_3,T_2)\frac{L_2(0)}{L_2(t)},
\end{equation*}
it can be seen that $U_1(\xi_1,T_1)$, $U_2(\xi_2,T_2)$ and $U_3(\xi_3,T_2)$ satisfy
\begin{align*}
\frac{\partial U_1(\xi_1,T_1)}{\partial T_1} &= D_1\frac{\partial^2U_1(\xi_1,T_1)}{\partial \xi_1^2}, \ -L_1(0) < \xi_1 < L_1(0), \\
\frac{\partial U_2(\xi_2,T_2)}{\partial T_2} &= D_2 \frac{\partial^2U_2(\xi_2,T_2)}{\partial \xi_2^2}, \ -L_2(0)-L_1(0) < \xi_2 < -L_1(0), \\
\frac{\partial U_3(\xi_3,T_2)}{\partial T_2} &= D_2 \frac{\partial^2U_3(\xi_3,T_2)}{\partial \xi_3^2}, \ L_1(0) < \xi_3 < L_1(0)+L_2(0), 
\end{align*}
which is the linear diffusion equation on multiple fixed domains. As for the previous multiple domain solutions, we impose the conservation of diffusive flux across the internal boundaries, noting that the velocity fields at the boundaries are continuous. We assume that there is a Dirac delta initial condition at $x = x_0$ where $-L_1(0) < x_0 < L_1(0)$ and hence $C_1(x,t)$ will have components arising from the initial mass and the re-introduced boundary flux, as before, while $C_2(x,t)$ and $C_3(x,t)$ will only have components from the internal boundary flux. Due to the growing domains, the weights for the boundary flux terms now depend on the transformations of $t$ and $C_i(x,t)$. We again require the weights sum to one and that, if $D_1 = D_2$, then $C_1(L_1,t) = C_3(L_1,t)$ due to the continuity of the velocity fields. The weights are
\begin{align*}
w_1^+(t,\tau) &= \frac{\sqrt{D_2}\sqrt{T_1(t,\tau)}L_2(\tau)/L_2(t)}{\sqrt{D_2}\sqrt{T_1(t,\tau)}L_2(\tau)/L_2(t)+\sqrt{D_1}\sqrt{T_2(t,\tau)}L_1(\tau)/L_1(t)}, \\ w_3^-(t,\tau) &= \frac{\sqrt{D_1}\sqrt{T_2(t,\tau)}L_1(\tau)/L_1(t)}{\sqrt{D_2}\sqrt{T_1(t,\tau)}L_2(\tau)/L_2(t)+\sqrt{D_1}\sqrt{T_2(t,\tau)}L_1(\tau)/L_1(t)},
\end{align*}
for the boundary at $x = L_1(t)$, and
\begin{align*}
w_1^-(t,\tau) &= \frac{\sqrt{D_2}\sqrt{T_1(t,\tau)}L_2(\tau)/L_2(t)}{\sqrt{D_2}\sqrt{T_1(t,\tau)}L_2(\tau)/L_2(t)+\sqrt{D_1}\sqrt{T_2(t,\tau)}L_1(\tau)/L_1(t)}, \\ w_2^+(t,\tau) &= \frac{\sqrt{D_1}\sqrt{T_2(t,\tau)}L_1(\tau)/L_1(t)}{\sqrt{D_2}\sqrt{T_1(t,\tau)}L_2(\tau)/L_2(t)+\sqrt{D_1}\sqrt{T_2(t,\tau)}L_1(\tau)/L_1(t)},
\end{align*}
for the boundary at $x = -L_1(t)$, where
\begin{equation*}
T_1(t,\tau) = \int_\tau^t \bigg(\frac{L_1(\tau)}{L_1(s)}\bigg)^2 \ \text{d}s, \qquad T_2(t,\tau) = \int_\tau^t \bigg(\frac{L_2(\tau)}{L_2(s)}\bigg)^2 \ \text{d}s.
\end{equation*}
The solutions are therefore
\begin{equation}
C_1(x,t) = H(x,t) + \int_0^t w_1^-(t,\tau)F_1^-(\tau)P_1^-(x,-L_1(t),t-\tau) + w_1^+(t,\tau)F_1^+(\tau)P_1^+(x,L_1(t),t-\tau) \ \text{d}\tau, \label{Eq:MultipleGrowingDomainsSolution1}
\end{equation}
where
\begin{align}
H(x,t) = &\frac{1}{\sqrt{4\pi D_1T_1(t)}}\frac{L_1(0)}{L_1(t)}\Bigg[\exp\bigg(-\frac{\Big(x\frac{L_1(0)}{L_1(t)}-x_0\Big)^2}{4D_1T_1(t)}\bigg)\nonumber \\ &+  \Bigg[\sum_{j = 1}^{\infty}(-1)^j  \exp\bigg(-\frac{\Big(x\frac{L_1(0)}{L_1(t)}-(-2jL_1(0)+(-1)^jx_0)\Big)^2}{4D_1T_1(t)}\bigg) \nonumber \\ &+ (-1)^j \exp\bigg(-\frac{\Big(x\frac{L_1(0)}{L_1(t)}-(2jL_1(0)+(-1)^jx_0)\Big)^2}{4D_1T_1(t)}\bigg)\Bigg], \label{Eq:MultipleGrowingDomainsSolution2}
\end{align}
and 
\begin{equation*}
F_1^-(t) = -D_1\frac{\partial H}{\partial x}\bigg|_{x=-L_1(t)}, \qquad F_1^+(t) = -D_1\frac{\partial H}{\partial x}\bigg|_{x=L_1(t)}.
\end{equation*}
The solution for domains two and three are
\begin{align}
C_2(x,t) &= \int_0^t w_2^+(t,\tau)F_1^-(\tau)P_2^+(x,-L_1(t),t-\tau)  \ \text{d}\tau, \label{Eq:MultipleGrowingDomainsSolution3} \\ 
C_3(x,t) &= \int_0^t w_3^-(t,\tau)F_1^+(\tau)P_3^-(x,L_1(t),t-\tau)  \ \text{d}\tau, \label{Eq:MultipleGrowingDomainsSolution4}
\end{align}
where $P_i^k(x,x_0,t)$ is defined as previously, albeit for the $i$th growing domain, rather than a fixed domain. \\

\subsection{Splitting and survival probabilities}
The proportion of the individuals that have yet to cross either external boundary, known as the survival probability, is a key metric for diffusive processes on growing domains \cite{simpson2015b,simpson2015c,yuste2016}. The proportion of the individuals that have crossed either the left or right external boundary, known as the left and right splitting probabilities, is similarly of interest. We can leverage the derived exact solutions to calculate these metrics, both for the single growing domain and multiple growing domain cases. Critically, we can obtain expressions for the time-varying splitting and survival probabilities, rather than just in the long-time limit. \\

For the case with a single growing domain, it is relatively straightforward to calculate the splitting and survival probabilities. The time-varying splitting probabilities for the boundaries at $x = -L(t)$ and $x = L(t)$, denoted $S^-(t)$ and $S^+(t)$, respectively, are simply the time integral of the appropriate boundary flux. The fluxes for the two boundaries are
\begin{align*}
F^-(t) = -D_1 \frac{\partial C(x,t)}{\partial x}\bigg|_{x=-L(t)} = &\sum_{j=0}^{\infty}(-1)^j \frac{(2j+1)L(0)+(-1)^jx_0}{\sqrt{4\pi D_1T(t)^3}}\bigg(\frac{L(0)}{L(t)}\bigg)^2 \\ &\times \exp\bigg(-\frac{((2j+1)L(0)+(-1)^jx_0)^2}{4D_1T(t)}\bigg), \\
F^+(t) = -D_1 \frac{\partial C(x,t)}{\partial x}\bigg|_{x=L(t)} = &\sum_{j=0}^{\infty} (-1)^j \frac{(2j+1)L(0)-(-1)^jx_0}{\sqrt{4\pi D_1T(t)^3}}\bigg(\frac{L(0)}{L(t)}\bigg)^2\\ &\times \exp\bigg(-\frac{((2j+1)L(0)-(-1)^jx_0)^2}{4D_1T(t)}\bigg).
\end{align*}
The time-varying splitting probabilities are therefore
\begin{align}
S^-(t) = \int_0^t F_1^-(\tau) \ \text{d}\tau &= \sum_{j=0}^{\infty} (-1)^j \text{erfc}\bigg(\frac{(2j+1)L(0)+(-1)^jx_0}{\sqrt{4 D_1T(t)}}\bigg), \label{Eq:LeftSplitSingleDomain} \\
S^+(t) = \int_0^t F_1^+(\tau) \ \text{d}\tau &= \sum_{j=0}^{\infty} (-1)^j \text{erfc}\bigg(\frac{(2j+1)L(0)-(-1)^jx_0}{\sqrt{4 D_1T(t)}}\bigg). \label{Eq:RightSplitSingleDomain}
\end{align}
The survival probability, $S(t)$, is simply the proportion of the population that has not crossed either boundary
\begin{align}
S(t) = 1 - S^-(t) - S^+(t) = &1 - \sum_{j=0}^{\infty}\Bigg[ (-1)^j \text{erfc}\bigg(\frac{(2j+1)L(0)+(-1)^jx_0}{\sqrt{4 D_1T(t)}}\bigg)\nonumber \\ &+ (-1)^j \text{erfc}\bigg(\frac{(2j+1)L(0)-(-1)^jx_0}{\sqrt{4 D_1T(t)}}\bigg)\Bigg]. \label{Eq:SurvivalSingleDomain}
\end{align}

For the case with multiple growing domains, the process to calculate the splitting and survival probabilities is more complicated. For an individual to cross an external boundary, it must first cross an internal boundary. We therefore first calculate the internal splitting probabilities, $S_1^-(t)$ and $S_1^+(t)$. Note that this is the probability that an individual crosses an internal boundary and does not re-enter the internal domain. This is calculated by weighting the internal boundary flux as before and taking the integral over time
\begin{equation*}
S_1^-(t) = \int_0^t w_2^+(t,\tau) F_1^-(\tau) \text{d}\tau, \qquad S_1^+(t) = \int_0^t w_3^-(t,\tau) F_1^+(\tau) \text{d}\tau.
\end{equation*}
A closed-form expression for the internal splitting probabilities can be obtained if $\sqrt{T_1(t,\tau)}L_1(t)/L_1(\tau) = \sqrt{T_2(t,\tau)}L_2(t)/L_2(\tau)$ holds for all $t$ and $\tau$, that is, that the ratio of the new domain size to the original domain size is the same for both the internal and external domains. Where this holds, the weights reduce to a constant, and hence
\begin{align*}
S_1^-(t) &= \frac{\sqrt{D_1}}{\sqrt{D_1}+\sqrt{D_2}} \sum_{j=0}^{\infty} (-1)^{j}\text{erfc}\bigg(\frac{(2j+1)L_1(0)+(-1)^jx_0}{\sqrt{4D_1T(t)}}\bigg), \\
S_1^+(t) &= \frac{\sqrt{D_1}}{\sqrt{D_1}+\sqrt{D_2}} \sum_{j=0}^{\infty} (-1)^{j}\text{erfc}\bigg(\frac{(2j+1)L_1(0)-(-1)^jx_0}{\sqrt{4D_1T(t)}}\bigg).
\end{align*}
If this condition does not hold, we can calculate the splitting probabilities via numerical integration. As before, the internal survival probability can be calculated according to $S_1(t) = 1 - S_1^-(t) - S_1^+(t)$. \\

To calculate the external splitting probabilities, that is, the probability that an individual crosses the boundary at either $x = -L_1(t) - L_2(t)$ or $x = L_1(t) + L_2(t)$, we first require an expression for the flux through the appropriate boundary. Recall that the solutions in domains two and three correspond to an integral over solution kernels that represent mass initially located at $x = \pm L_1(t)$. The flux through the boundary at $x = L_1(t)+L_2(t)$ at time $t$ due to mass that is placed at $x = L_1(t)$ at $t = \tau$ is
\begin{align*}
F_3^+(t,\tau) = &-D_2\frac{\partial P_3(x,L_1(t),t-\tau)}{\partial x}\bigg|_{x=L_1(t)+L_2(t)} \\ & = \sum_{j=0}^{\infty} (-1)^j \frac{(2j+1)(L_1(\tau)+L_2(\tau))-(-1)^jL_1(\tau)}{\sqrt{\pi D_2T_2(t,\tau)^3}}\bigg(\frac{L_2(\tau)}{L_2(t)}\bigg)^2 \times \\ &\exp\bigg(-\frac{\Big((2j+1)(L_1(\tau) + L_2(\tau))-(-1)^jL_1(\tau)\Big)^2}{4D_2T_2(t,\tau)}\bigg).
\end{align*}
The survival probability at time $t$ in domain three for the mass that is placed at $x = L_1(t)$ at $t = \tau$ is obtained through the relationship 
\begin{equation*}
1-S_3(t,\tau) = \int_\tau^t F_3^+(s,\tau) \ \text{d}s = \sum_{j=0}^{\infty} (-1)^j\text{erfc}\bigg(\frac{(2j+1)(L_1(\tau) + L_2(\tau))-(-1)^jL_1(\tau)}{\sqrt{4D_2T_2(t,\tau)}}\bigg),
\end{equation*}
noting that there is no flux from domain three into domain one, as can be seen from the form of the solution kernel. While agents in the simulation do cross the internal boundaries multiple times, these are accounted for in the kernel via the weighting of the flux. These results are for a solution kernel corresponding to a mass that is placed in a single instant; however the solution in domain three involves an integral over the mass placed at each instant. Therefore, the overall splitting probability for the boundary at $x = L_1(t) + L_2(t)$ is
\begin{equation*}
S^+(t) = \int_0^t F_1^+(\phi) \int_\phi^t w_3^-(s,\phi) F_3^+(s,\phi) \ \text{d}s \ \text{d}\phi.
\end{equation*}
The three components in the integral are the flux across the internal boundary ($F_1^+(\phi)$), the fraction of that flux that is placed into domain three ($w_3^-(s,\phi)$) and the flux across the external boundary ($F_3^+(s,\phi)$), given that there was mass placed at $x = L_1(\phi)$. In the case where $\sqrt{T_1(t,\tau)}L_1(t)/L_1(\tau) = \sqrt{T_2(t,\tau)}L_2(t)/L_2(\tau)$ we can use the expressions for the internal boundary flux and the survival probability for domain three to obtain a single integral expression for the splitting probability for the boundary at $x = L_1(t) + L_2(t)$ 
\begin{align}
S^+(t) = & \frac{\sqrt{D_1}}{\sqrt{D_1}+\sqrt{D_2}} \int_0^t \bigg[\sum_{j=0}^{\infty} (-1)^j \frac{(2j+1)L_1(0)-(-1)^jx_0}{\sqrt{4\pi D_1T_1(\phi)^3}}\bigg(\frac{L_1(0)}{L(\phi)}\bigg)^2 \nonumber \\ &\times \exp\bigg(-\frac{((2j+1)L_1(0)-(-1)^jx_0)^2}{4D_1T_1(\phi)}\bigg)\bigg] \times \nonumber \\ &\bigg[\sum_{j=0}^{\infty} (-1)^j\text{erfc}\bigg(\frac{(2j+1)(L_1(\phi) + L_2(\phi))-(-1)^jL_1(\phi)}{\sqrt{4D_2T_2(t,\phi)}}\bigg)\bigg] \ \text{d}\phi. \label{Eq:RightSplitMultipleDomains}
\end{align}
As before, if this relationship does not hold, we can use numerical techniques to evaluate the integral. Following similar arguments, we obtain the splitting probability for the boundary at $x = -L_1(t) - L_2(t)$
\begin{equation*}
S^-(t) = \int_0^t F_1^-(\phi) \int_\phi^t w_2^+(s,\phi) F_2^-(s,\phi) \ \text{d}s \ \text{d}\phi,
\end{equation*}
which, provided $\sqrt{T_1(t,\tau)}L_1(t)/L_1(\tau) = \sqrt{T_2(t,\tau)}L_2(t)/L_2(\tau)$ is satisfied, can be expressed
\begin{align}
S^-(t) = & \frac{\sqrt{D_1}}{\sqrt{D_1}+\sqrt{D_2}} \int_0^t \bigg[\sum_{j=0}^{\infty} (-1)^j \frac{(2j+1)L_1(0)+(-1)^jx_0}{\sqrt{4\pi D_1T_1(\phi)^3}}\bigg(\frac{L_1(0)}{L(\phi)}\bigg)^2 \nonumber \\ &\times \exp\bigg(-\frac{((2j+1)L_1(0)+(-1)^jx_0)^2}{4D_1T_1(\phi)}\bigg)\bigg] \times \nonumber \\ &\bigg[\sum_{j=0}^{\infty} (-1)^j\text{erfc}\bigg(\frac{(2j+1)(L_1(\phi)+ L_2(\phi))-(-1)^jL_1(\phi)}{\sqrt{4D_2T_2(t,\phi)}}\bigg)\bigg] \ \text{d}\phi. \label{Eq:LeftSplitMultipleDomains}
\end{align}
The overall survival probability (i.e. that an agent remains in any domain) is given by
\begin{equation}
S(t) = 1 - S^-(t) - S^+(t). \label{Eq:SurvivalMultipleDomains}
\end{equation}

\section{Results}

For all results we choose unit lattice widths and timesteps such that $\Delta = \delta t = 1$. We consider three functional forms for domain \textcolor{black}{evolution}:
\begin{itemize}
\item \textit{Linear growth}. Here an individual domain undergoes (positive) growth that is linear in time according to
\begin{equation}
L_i(t) = L_i(0) + \beta_it, \label{Eq:LinearDomainGrowth}
\end{equation}
where $\beta_i > 0$. The corresponding time transformation is
\begin{equation}
T_i(t,\tau) = \bigg(\frac{L_i(0)+\beta_i \tau}{L_i(0)+\beta_i t}\bigg)(t-\tau). \label{Eq:LinearTimeTransformation}
\end{equation}
Note that the time transformation $T_i(t)$ is obtained by setting $\tau = 0$. It is possible to select $\beta_i < 0$; however, this requires the restriction that $t < -L_i(0)/\beta_i$ to ensure that the domain size remains positive and the solution does not experience finite-time blow-up.
\item \textit{Exponential growth}. Here an individual domain undergoes growth that is exponential in time according to
\begin{equation}
L_i(t) = L_{i,\text{min}} + (L_i(0)-L_{i,\text{min}})\exp(-\beta_i t). \label{Eq:ExpDomainGrowth}
\end{equation}
For exponential growth we do not have a restriction on the sign of $\beta_i$ as the domain will approach the finite size $L_{i,\text{min}} > 0$ when $\beta_i > 0$, and hence we can study examples of both positive and negative growth. The corresponding time transformation is 
\begin{align}
T_i(t,\tau) =& \frac{\Big[\exp(-\beta_i\tau)\big(L_i(0)-L_{i,\text{min}}\big)+L_{i,\text{min}}\Big]^2}{\beta_i L_{i,\text{min}}^2} \nonumber \\ \times  & \bigg[ \big(L_i(0)-L_{i,\text{min}}\big)\bigg(\frac{1}{L_{i,\text{min}}\big[1-\exp(\beta_i\tau)]-L_i(0)} + \frac{1}{L_i(0)+L_{i,\text{min}}\big[\exp(\beta_i t)-1\big]}\bigg) \nonumber \\ & - \log\Big(L_i(0)+L_{i,\text{min}}\big[\exp(\beta_i \tau)-1\big]\Big) + \log\Big(L_i(0) + L_{i,\text{min}}\big[\exp(\beta_i t)-1\big]\Big)\bigg]. \label{Eq:ExpTimeTransformation}
\end{align}
\item \textit{Oscillatory \textcolor{black}{evolution}}. Here an individual domain undergoes \textcolor{black}{evolution} that oscillates in time according to
\begin{equation}
L_i(t) = L_i(0) + (L_i(0) - L_{i,\text{min}})\sin(\beta_i t) \label{Eq:SinDomainGrowth}.
\end{equation}
Again, for oscillatory \textcolor{black}{evolution}, there is no restriction on the sign of $\beta_i$ as this simply dictates whether the domain initially experiences positive or negative growth. The domain size is bounded according to $L_{i,\text{min}} \leq L_i(t) \leq 2L_i(0)-L_{i,\text{min}}$, where $0 < L_{i,\text{min}} \leq L_i(0)$. The corresponding time transformation is

\begin{align}
T_i(t,\tau) =& \Bigg(\frac{L_i(0)+\big(L_i(0)-L_{i,\text{min}}\big)\sin(\beta_i \tau)}{L_i(0)+(L_i(0)-L_{i,\text{min}})\sin(\beta_i t)}\Bigg) \nonumber \\ &\times \Bigg(\frac{1}{\beta_iL_{i,\text{min}}\big[2L_i(0)-L_{i,\text{min}}\big]\big[L_i(0)^2-(L_i(0)-L_{i,\text{min}})^2\big]^{1/2}}\Bigg) \nonumber \\ &\times \Bigg[\big[L_i(0)-L_{i,\text{min}}\big]\big[L_i(0)^2-(L_i(0)-L_{i,\text{min}})^2\big]^{1/2}\nonumber \\ & \times \Big(\cos(\beta_i t)\big[L_i(0) + (L_i(0)-L_{i,\text{min}})\sin(\beta_i \tau)\big] \nonumber \\ & - \cos(\beta_i \tau)\big[L_i(0) + (L_i(0)-L_{i,\text{min}})\sin(\beta_i t)\big]\Big) \nonumber \\ & + 2L_i(0)\big[L_i(0)+(L_i(0)-L_{i,\text{min}})\sin(\beta_i \tau)\big]\big[L_i(0)+(L_i(0)-L_{i,\text{min}})\sin(\beta_i t)\big] \nonumber  \\ & \times \Bigg[ \tan^{-1}\Bigg(\frac{L_i(0)\Big[\tan\big(\beta_i t/2)+1\Big]-L_{i,\text{min}}}{\big[L_i(0)^2-(L_i(0)-L_{i,\text{min}})^2\big]^{1/2}}\Bigg) \nonumber \\ &- \tan^{-1}\Bigg(\frac{L_i(0)\Big[\tan\big(\beta_i \tau/2)+1\Big]-L_{i,\text{min}}}{\big[L_i(0)^2-(L_i(0)-L_{i,\text{min}})^2\big]^{1/2}}\Bigg)\Bigg]\Bigg] \nonumber \\ &+
\Big[L_i(0)+(L_i(0)- L_{i,\text{min}})\sin\big(\beta_i\tau\big)\Big]^2\Bigg[\frac{2\pi L_i(0)}{\beta_i\big[L_i(0)^2-(L_i(0)-L_{i,\text{min}})^2\big]^{3/2}}\Bigg] \nonumber \\ &\times \Bigg[\bigg\lfloor\frac{\beta_i t + \pi}{2\pi}\bigg\rfloor - \bigg\lfloor\frac{\beta_i \tau + \pi}{2\pi}\bigg\rfloor\Bigg]. \label{Eq:SinTimeTransformation}
\end{align}
\end{itemize}

To ensure that our solutions are consistent with those derived in previous investigations \cite{simpson2015c}, we compare the average individual density in the random walk against the exact solution for a single fixed domain and a single growing domain (Appendix A, Figures 9 and 10).  As in previous investigations \cite{simpson2015c}, we observe a close match between the density profiles obtained from repeated realisations of the random walk and the exact solutions. \textcolor{black}{A formal derivation of a continuum model from a lattice-based random walk on a growing domain can be found in \cite{baker2010}.} We also examine whether the derived exact solutions for multiple fixed domains are consistent with the average behaviour in the random walk. We consider both the case where $D_1 = D_2$ and where $D_1 \neq D_2$ (Appendix A, Figures 11 and 12). In both cases we observe that the exact solutions match the average random walk behaviour well, which suggests that the derived solutions are valid. \\

\begin{figure}
\begin{center}
\includegraphics[width=1.0\textwidth]{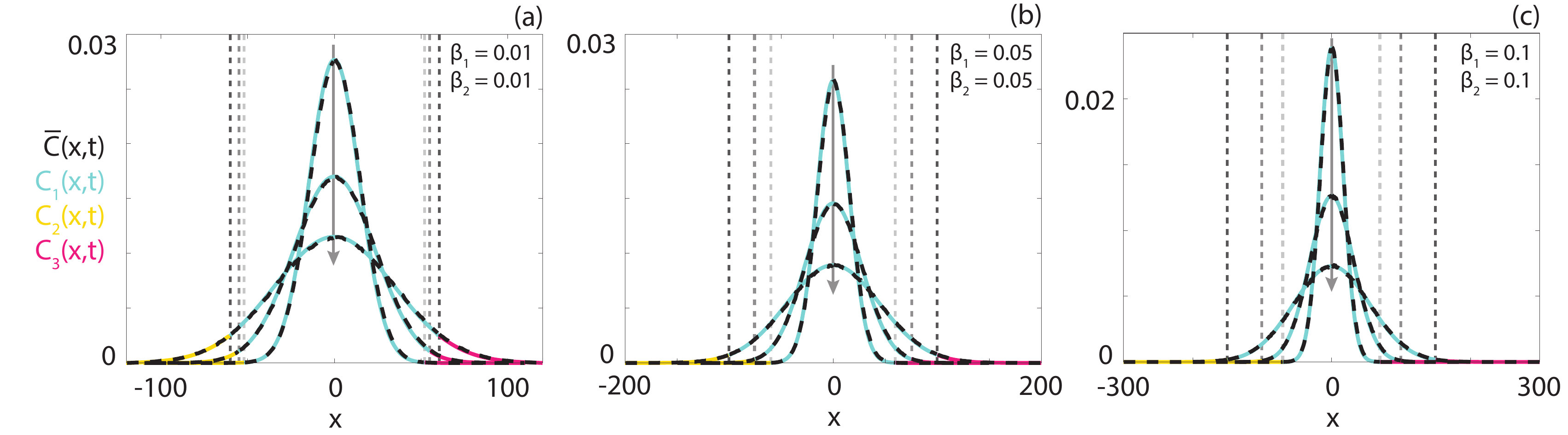}
\caption{Comparison between the average behaviour in the lattice-based random walk $\overline{C}(x,t)$ (black, dashed) and the exact solutions $C_1(x,t)$ (cyan), $C_2(x,t)$ (orange) and $C_3(x,t)$ (pink) as defined in Equations \eqref{Eq:MultipleGrowingDomainsSolution1}-\eqref{Eq:MultipleGrowingDomainsSolution4} for multiple linearly growing domains (Equations \eqref{Eq:LinearDomainGrowth}-\eqref{Eq:LinearTimeTransformation}) and domain-independent diffusivities. Parameters used are $D_1 = D_2 = 0.5$, $L_1(0) = L_2(0) = 50$, $N = 1000$, $x_0 = 0$, (a) $\beta_1 = \beta_2 = 0.01$, (b) $\beta_1 = \beta_2 = 0.05$, (c) $\beta_1 = \beta_2 = 0.1$. Solution profiles are presented at $t = 200$, $t = 500$ and $t = 1000$.  The arrow indicates the direction of increasing time. Dashed grey lines correspond to the position of the boundary. Average random walk behaviour is obtained from 5000 identically-prepared realisations of the random walk.}
\label{F3}
\end{center}
\end{figure}

We now impose domain growth and examine whether the derived exact solutions are valid for the case with multiple growing domains. As before, we verify the solutions via comparison against the average random walk behaviour. We first consider examples with positive linear domain growth for three different rates of growth and present the results in Figure \ref{F3}. In each case we observe that the exact solutions match the average random walk behaviour in each of the three domains. The exact solutions for $C_1(x,t)$, $C_2(x,t)$ and $C_3(x,t)$ are shown in cyan, orange and pink, respectively. The location of the internal boundary is shown via the dashed grey line, which increases in intensity as time increases. We observe that as we increase the rate of domain growth a smaller proportion of individuals leaves the inner domain. We note that as $D_1 = D_2$ and the ratio $L_1(t)/L_1(0) = L_2(t)/L_2(0) = L_3(t)/L_3(0)$ is consistent between domains this problem is comparable to the problem of domain growth on a single domain. \\

\begin{figure}
\begin{center}
\includegraphics[width=1.0\textwidth]{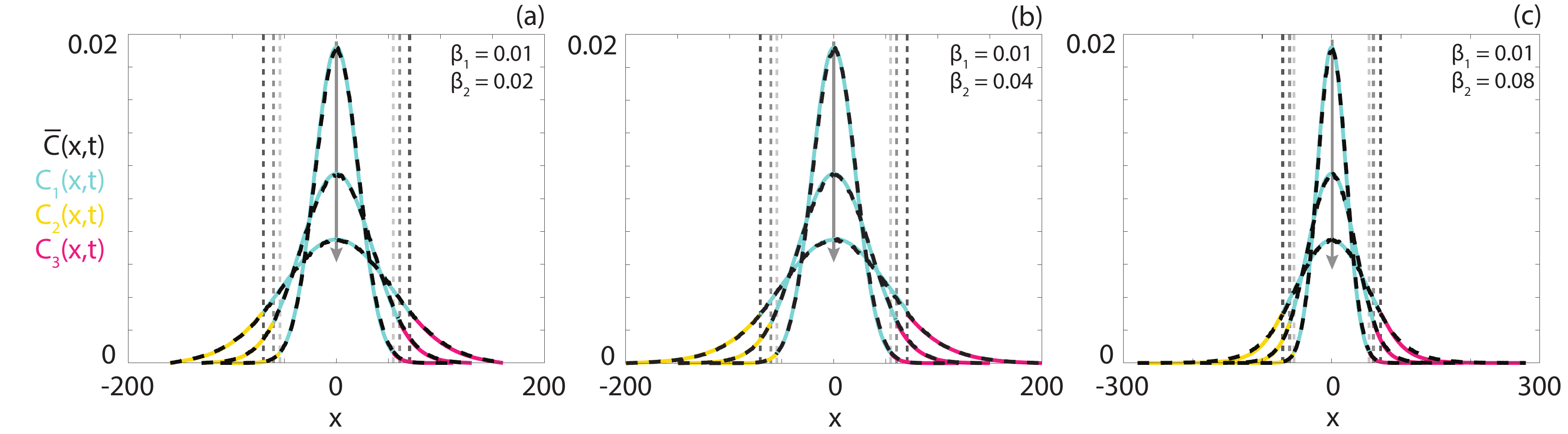}
\caption{Comparison between the average behaviour in the lattice-based random walk $\overline{C}(x,t)$ (black, dashed) and the exact solutions $C_1(x,t)$ (cyan), $C_2(x,t)$ (orange) and $C_3(x,t)$ (pink) as defined in Equations \eqref{Eq:MultipleGrowingDomainsSolution1}-\eqref{Eq:MultipleGrowingDomainsSolution4} for multiple linearly growing domains (Equations \eqref{Eq:LinearDomainGrowth}-\eqref{Eq:LinearTimeTransformation}) and domain-independent diffusivities. Parameters used are $D_1 = D_2 = 0.5$, $L_1(0) = L_2(0) = 50$, $N = 1000$, $x_0 = 0$, (a) $\beta_1 = 0.01$, $\beta_2 = 0.02$, (b) $\beta_1 = 0.01$,  $\beta_2 = 0.04$, (c) $\beta_1 = 0.01$, $\beta_2 = 0.08$. Solution profiles are presented at $t = 400$, $t =1000$ and $t = 2000$. The arrow indicates the direction of increasing time. Dashed grey lines correspond to the position of the boundary. Average random walk behaviour is obtained from 5000 identically-prepared realisations of the random walk.}
\label{F4}
\end{center}
\end{figure}

If the relative rates of domain growth are different between domains, we expect different behaviour to the single domain problem. We consider three examples where $D_1 = D_2$ and $\beta_1 \neq \beta_2$ and present solution profiles obtained from repeated realisations of the random walk and enumerating the exact solutions for three different domain growth rate pairs in Figure \ref{F4}. In each example $\beta_1 = 0.01$, while $\beta_2 = 0.02$ in Figure \ref{F4}(a), $\beta_2 = 0.04$ in Figure \ref{F4}(b) and $\beta_2 = 0.08$ in Figure \ref{F4}(c). The exact solutions match the average random walk behaviour well. As $\beta_2$ increases, we see fewer individuals reaching the external boundaries, though the solution in the inner domain is largely consistent. The additional domain growth serves to stretch the solution profile (via advection) in the outer domains. However, this additional advection is not sufficient for the individuals to reach the outer boundary at higher $\beta_2$ values. \\

\begin{figure}
\begin{center}
\includegraphics[width=0.5\textwidth]{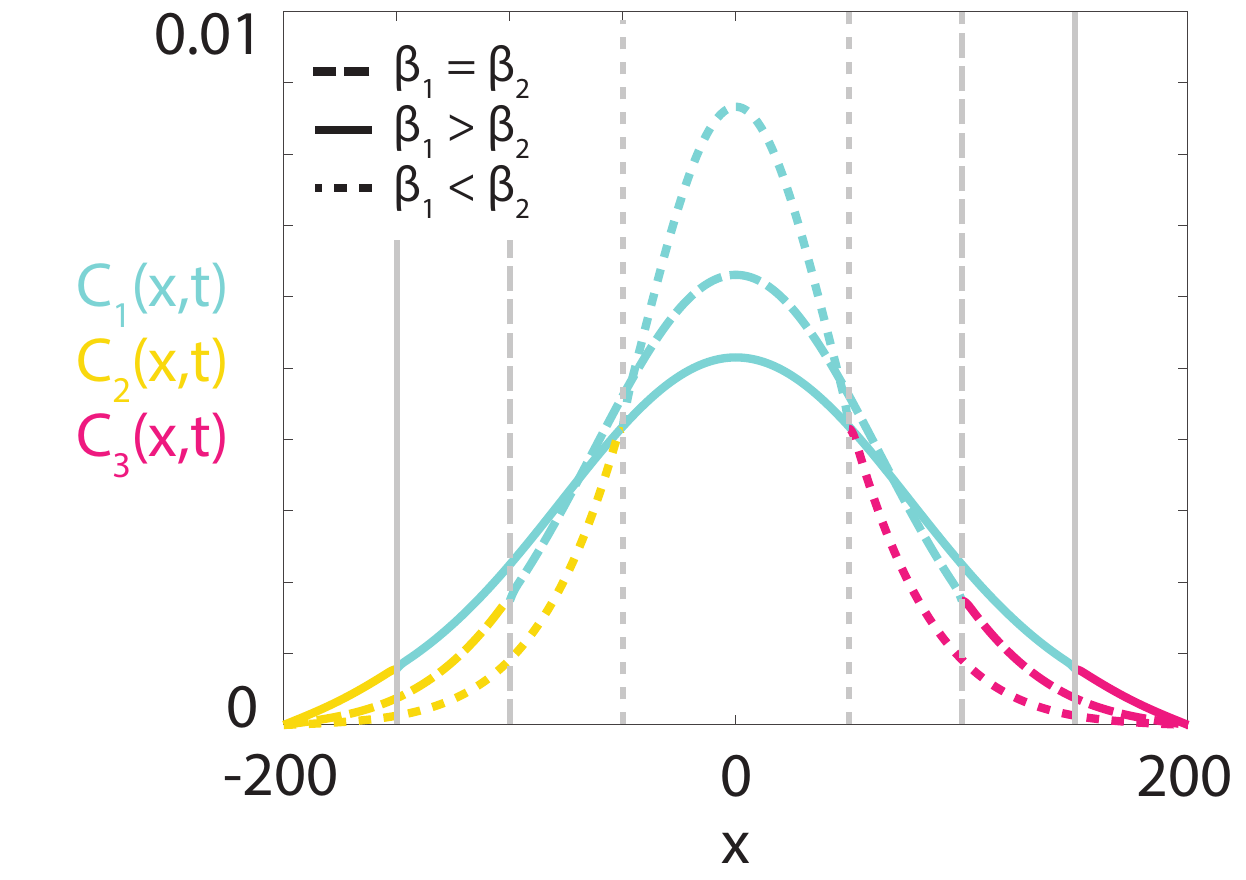}
\caption{\textcolor{black}{Comparison between heterogeneous and homogeneous domain growth. Exact solutions $C_1(x,t)$ (cyan), $C_2(x,t)$ (orange) and $C_3(x,t)$ (pink) as defined in Equations \eqref{Eq:MultipleGrowingDomainsSolution1}-\eqref{Eq:MultipleGrowingDomainsSolution4} for multiple linearly growing domains (Equations \eqref{Eq:LinearDomainGrowth}-\eqref{Eq:LinearTimeTransformation}) with domain-independent diffusivities. Parameters used are $D_1 = D_2 = 0.5$, $L_1(0) = L_2(0) = 50$, $\beta_1 = \beta_2 = 0.025$ (homogeneous growth, dashed lines), $\beta_1 = 0.05$, $\beta_2 = 0$ (heterogeneous growth, solid lines), $\beta_1 = 0$, $\beta_2 = 0.05$ (heterogeneous growth, dotted lines). Solution profiles are presented at $t = 2000$. Grey lines correspond to the position of the boundary.}}
\label{F5}
\end{center}
\end{figure}

\textcolor{black}{We highlight the impact of heterogeneous domain growth in Figure \ref{F5}. We consider the three possible relationships between $\beta_1$ and $\beta_2$ to investigate the differences between homogeneous and heterogeneous domain growth. That is, we examine: (i) $\beta_1 = \beta_2$ (homogeneous growth); (ii) $\beta_1 > \beta_2$ (heterogeneous growth), and; (iii) $\beta_1 < \beta_2$ (heterogeneous growth), noting that we ensure the total amount of domain growth is consistent across all three examples. These results illustrate a marked difference between homogeneous and heterogeneous domain growth. If the heterogeneity manifests in slower growth in the inner domain we see that the spread of the population is inhibited relative to the homogeneous growth case. In contrast, if the inner domain experiences more rapid growth then the population spreads faster than in the case of homogeneous domain growth. \\}

\begin{figure}
\begin{center}
\includegraphics[width=1.0\textwidth]{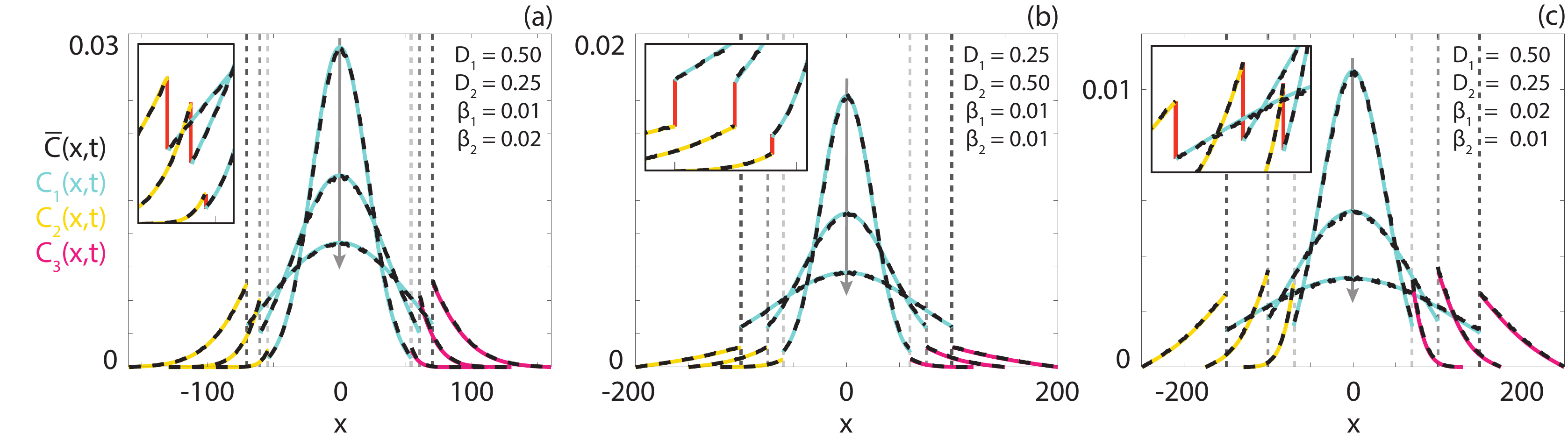}
\caption{Comparison between the average behaviour in the lattice-based random walk $\overline{C}(x,t)$ (black, dashed) and the exact solutions $C_1(x,t)$ (cyan), $C_2(x,t)$ (orange) and $C_3(x,t)$ (pink) as defined in Equations \eqref{Eq:MultipleGrowingDomainsSolution1}-\eqref{Eq:MultipleGrowingDomainsSolution4} for multiple linearly growing domains (Equations \eqref{Eq:LinearDomainGrowth}-\eqref{Eq:LinearTimeTransformation}) and domain-dependent diffusivities. \textcolor{black}{Inserts highlight the jump discontinuities (red).} Parameters used are $L_1(0) = L_2(0) = 50$, $N = 1000$, $x_0 = 0$, (a) $D_1 = 0.5$, $D_2 = 0.25$,  $\beta_1 = 0.01$, $\beta_2 = 0.02$, (b) $D_1 = 0.25$, $D_2 = 0.5$,  $\beta_1 = 0.01$, $\beta_2 = 0.01$, (c) $D_1 = 0.5$, $D_2 = 0.25$,  $\beta_1 = 0.02$, $\beta_2 = 0.01$. Solution profiles are presented at (a) $t = 400$, $t =1000$ and $t = 2000$, (b),(c) $t = 1000$, $t = 2000$ and $t = 5000$. The arrow indicates the direction of increasing time. Dashed grey lines correspond to the position of the boundary. Average random walk behaviour is obtained from 5000 identically-prepared realisations of the random walk.}
\label{F6}
\end{center}
\end{figure}

We now consider examples where $D_1 \neq D_2$. The form of the exact solution suggests that there will be a jump discontinuity at the internal boundary that satisfies $D_1C_1(L_1(t),t) = D_2C_3(L_1(t),t)$. \textcolor{black}{The presence of jump discontinuities in the density has an intuitive explanation, as the net movement between two sites that have different probabilities of movement is balanced by an equivalent (but proportionally opposite) difference in the number of agents occupying the sites.} For the three examples considered, presented in Figure \ref{F6}, we observe this jump discontinuity where if $D_1 > D_2$ then $C_1(L_1(t),t) < C_3(L_1(t),t)$ and if $D_1 < D_2$ then $C_1(L_1(t),t) > C_3(L_1(t),t)$, as expected. The jump discontinuity is accurately captured in both the exact solution and the average random walk behaviour, and we again see that the exact solution closely describes the average behaviour in the random walk. \textcolor{black}{A detailed discussion of similar jump discontinuities that arise across internal boundaries can be found in \cite{gudnason2018}.} \\

\begin{figure}
\begin{center}
\includegraphics[width=1.0\textwidth]{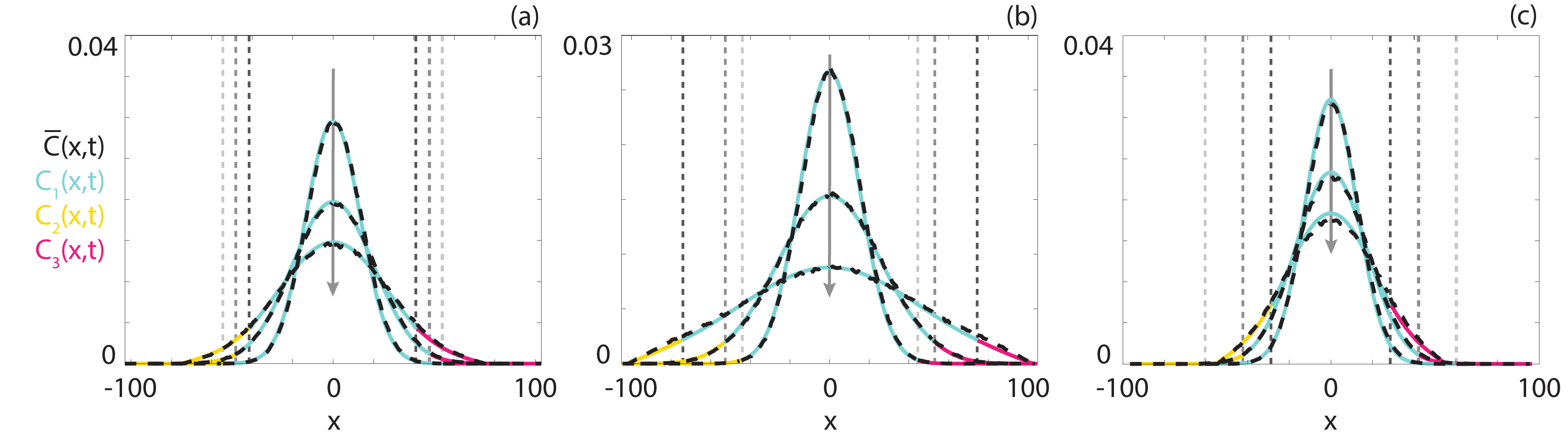}
\caption{Comparison between the average behaviour in the lattice-based random walk $\overline{C}(x,t)$ (black, dashed) and the exact solutions $C_1(x,t)$ (cyan), $C_2(x,t)$ (orange) and $C_3(x,t)$ (pink) as defined in Equations \eqref{Eq:MultipleGrowingDomainsSolution1}-\eqref{Eq:MultipleGrowingDomainsSolution4} for multiple exponentially growing domains (Equations \eqref{Eq:ExpDomainGrowth}-\eqref{Eq:ExpTimeTransformation}). Parameters used are $D_1 = D_2 = 0.5$, $N = 1000$, $x_0 = 0$, (a) $L_1(0) = L_2(0) = 60$, $L_{1,\text{min}} = L_{2,\text{min}} = 30$, $\beta_1 = 1\times10^{-3}$, $\beta_2 = 2\times10^{-3}$, (b) $L_1(0) = 40$, $L_2(0) = 80$, $L_{1,\text{min}} = L_{2,\text{min}} = 20$, $\beta_1 = -1\times10^{-3}$, $\beta_2 = 2\times10^{-3}$, (c) $L_1(0) = 80$, $L_2(0) = 40$, $L_{1,\text{min}} = L_{2,\text{min}} = 20$, $\beta_1 = 2\times10^{-3}$, $\beta_2 = -1\times10^{-3}$. Solution profiles are presented at $t = 200$, $t = 500$ and $t = 1000$. The arrow indicates the direction of increasing time. Dashed grey lines correspond to the position of the boundary. Average random walk behaviour is obtained from 5000 identically-prepared realisations of the random walk.}
\label{F7}
\end{center}
\end{figure}

Thus far, we have only considered examples where the domains have experienced linear growth. We now examine three cases of exponential growth, and present the average behaviour in the random walk and the profiles obtained from the exact solutions in Figure \ref{F7}. In Figure \ref{F7}(a), both domains decrease in size. In Figure \ref{F7}(b), the inner domain experiences positive growth while the outer domains experience negative growth. Finally, in Figure \ref{F7}(c), the inner domain undergoes negative growth and the outer domain undergoes positive growth. Critically, we select a positive value of $L_{i,\text{min}}$ so that even if the domain shrinks in size we do not experience finite-time blow-up in the solution due to the collision of the solution characteristics. In each case, there is close agreement between the average random walk behaviour and the exact solution, which confirms that our derived solutions are appropriate for describing negative domain growth. \\

\begin{figure}
\begin{center}
\includegraphics[width=1.0\textwidth]{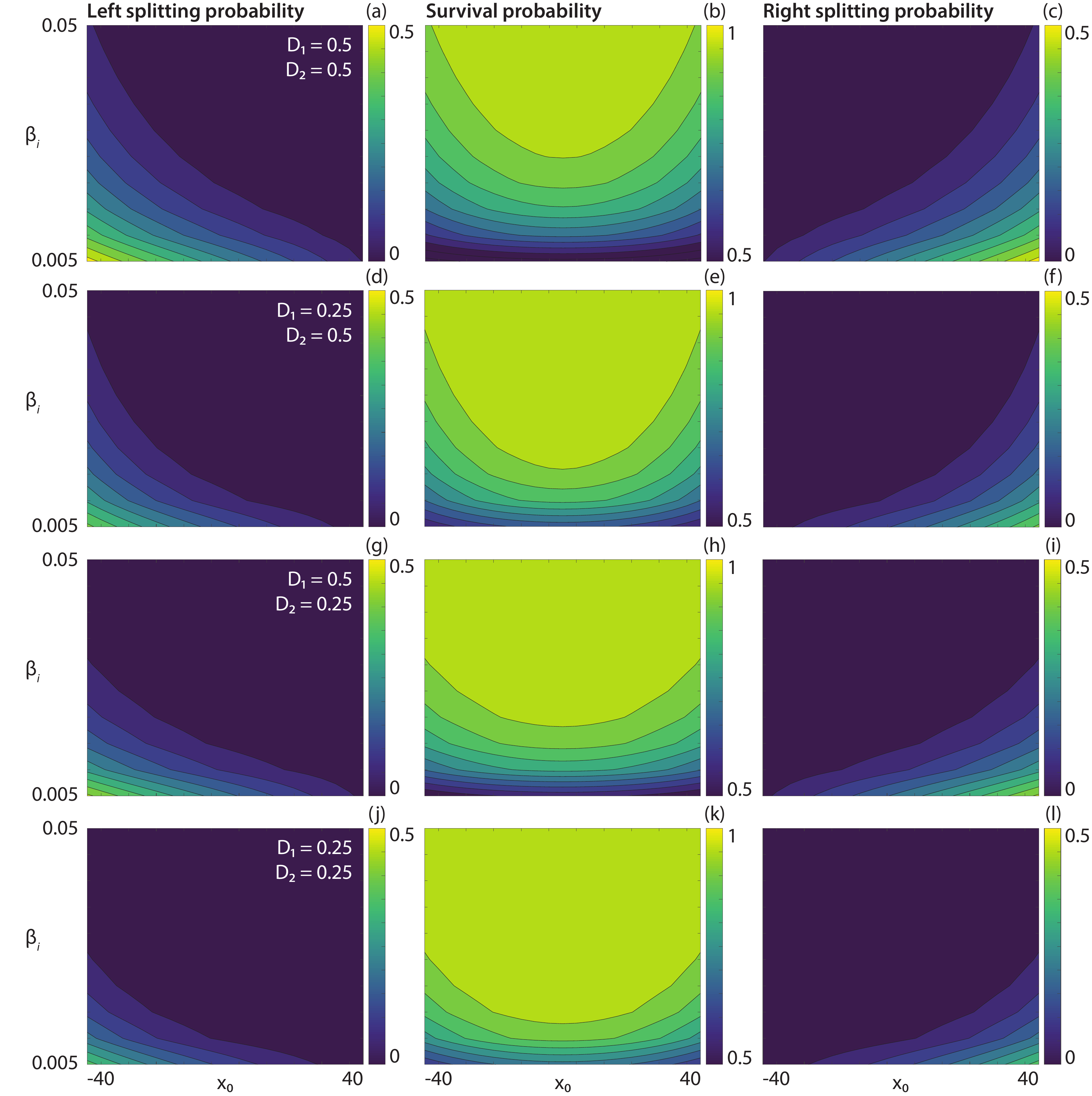}
\caption{(a),(d),(g),(j) Left splitting probability, (b),(e),(h),(k) survival probability and (c),(f),(j),(l) right splitting probability obtained from the exact solutions defined in Equations \eqref{Eq:RightSplitMultipleDomains}-\eqref{Eq:SurvivalMultipleDomains} for multiple linearly growing domains (Equations \eqref{Eq:LinearDomainGrowth}-\eqref{Eq:LinearTimeTransformation}) for a range of domain growth rates and initial locations. Parameter used are $L_1(0) = L_2(0) = 50$, (a)-(c) $D_1 = D_2 = 0.5$, (d)-(f) $D_1 = 0.25$, $D_2 = 0.5$, (g)-(i) $D_1 = 0.5$, $D_2 = 0.25$, (j)-(l) $D_1 = D_2 = 0.25$. Probabilities are presented at $t = 10^5$.}
\label{F8}
\end{center}
\end{figure}

One benefit of the exact solutions is that (near) steady state values of statistics of interest, such as the splitting and survival probabilities, can be enumerated without the need to perform simulations over long time periods. In Equations \eqref{Eq:LeftSplitSingleDomain}-\eqref{Eq:SurvivalMultipleDomains}, we derived expressions for the time-varying left and right splitting probabilities and survival probability. We verify that these probabilities are correct for both the single and multiple growing domains by comparing the derived expressions against the probabilities obtained from repeated realisations of the random walk (Appendix A, Figure 13). We highlight the usefulness of the expressions by exploring how domain growth rate, initial location and domain-dependent diffusivity influence the splitting and survival probabilities in Figure \ref{F8}. Uniformly increasing the diffusivity increases both the splitting probabilities while decreasing the survival probability. Interestingly, comparing the results where $D_1 = 0.25$ and $D_2 = 0.5$ (Figures \ref{F8}(d)-(f)) against the results where $D_1 = 0.5$ and $D_2 = 0.25$ (Figures \ref{F8}(g)-(i)) suggests that, in terms of an individual reaching the external boundary, an increased rate of movement in the outer domain is more important than in the inner domain. \\

\begin{figure}
\begin{center}
\includegraphics[width=0.8\textwidth]{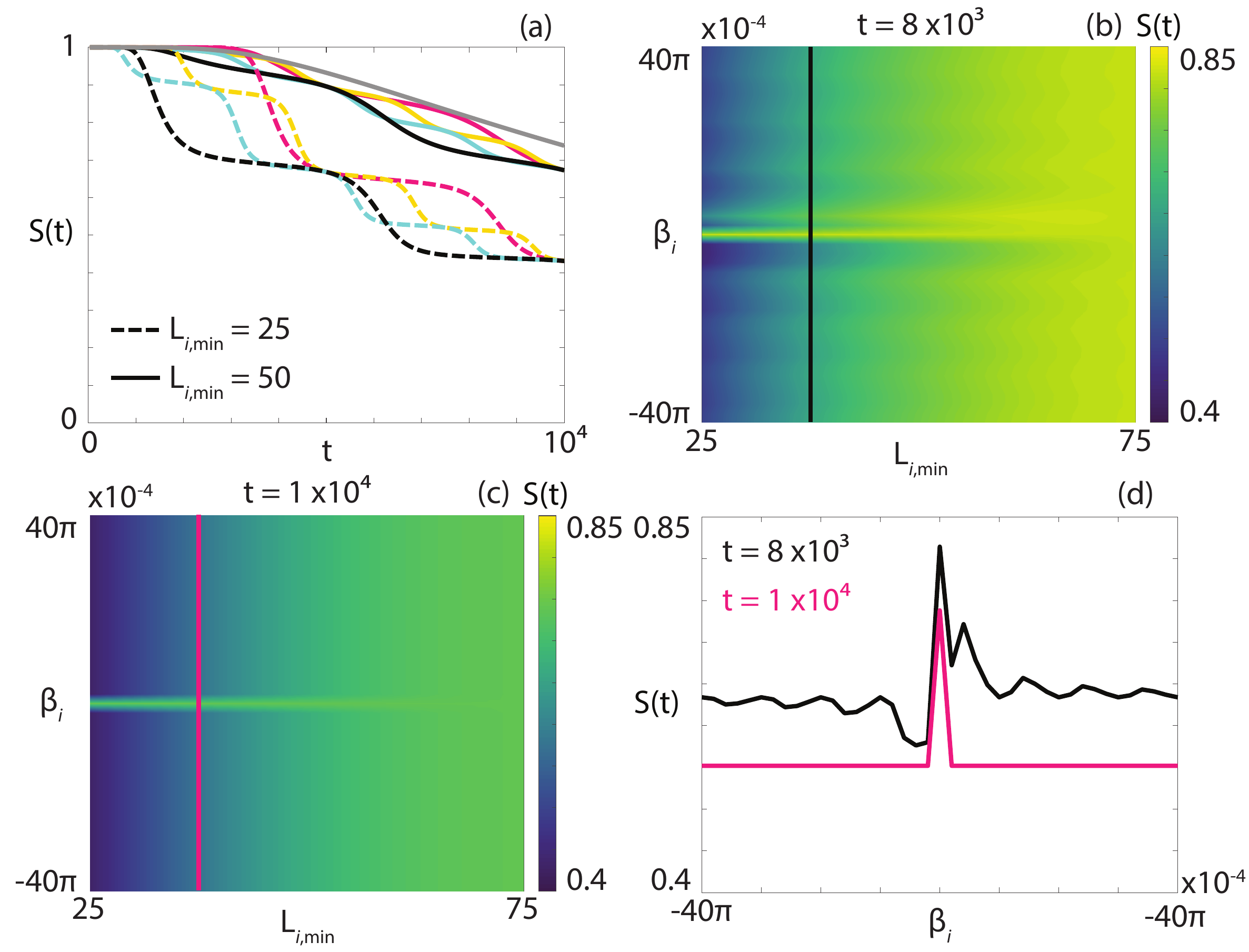}
\caption{(a) Survival probability over time as defined in Equations \eqref{Eq:RightSplitMultipleDomains}-\eqref{Eq:SurvivalMultipleDomains}  for multiple oscillating domains (Equations \eqref{Eq:SinDomainGrowth}-\eqref{Eq:SinTimeTransformation}). The grey line corresponds to no oscillation ($\beta_i = 0$). The solid lines correspond to $L_{i,\text{min}} = 50$. The dashed lines correspond to $L_{i,\text{min}} = 25$. Oscillation rates are $\beta_i = -8\pi \times 10^{-4}$ (cyan), $-4\pi \times 10^{-4}$ (black), $4\pi \times 10^{-4}$ (orange), $8\pi \times 10^{-4}$ (pink). (b),(c) Survival probability at (b) $t = 8\times10^3$ and (c) $t=1\times10^4$ for a range of oscillation rates and minimum domain sizes. (d) Survival probability at $t = 8\times10^3$ (black) and $t=1\times10^4$ (pink) for $L_{i,\text{min}} = 35$. The black line in (b) and the pink line in (c) correspond to the profiles in (d). Parameters used are $L_1(0) = L_2(0) = 75$, $D_1 = D_2 = 0.5$, and $x_0 = 0$.}
\label{F9}
\end{center}
\end{figure}

Finally, we consider the question of survival probabilities on oscillatory evolving domains. That is, we consider domains that oscillate between a minimum (but positive) and maximum size as defined in Equation \eqref{Eq:SinDomainGrowth}. Contractile cells, such as smooth muscle cells and cardiomyocytes, exhibit this behaviour \cite{bers2002}. Cardiomyocytes contract and relax with a regular rhythm as the heart beats, while various molecular species undergo diffusion within the cell \cite{bers2002}. We examine how the minimum domain size ($L_{i,\text{min}}$) and the rate of oscillation ($\beta_i$) impact the survival probability, and present the results in Figure \ref{F9}. We see that in all cases, after one oscillation has been completed, an individual is more likely to have reached the external boundary in comparison to the case where there is no oscillation (Figure \ref{F9}(a)). This is interesting as the average length of the domain is not altered by the presence of oscillation. However, as the size of individuals' movement events is independent of the overall size of the domain, a single movement event on a domain that is at its minimum size is proportionally larger than if at the domain is at its initial size. We see that for smaller minimum domain sizes the survival probability decreases more rapidly. We explore how changing the rates of oscillation and minimum domain sizes affects the survival probability at $t = 8\times10^3$ and $t = 1\times10^4$ (Figures \ref{F9}(b),(c)). The times and oscillation rates are chosen such that at $t = 1\times10^4$ all domains have completed an integer number of oscillations and are at the initial domain size. Interestingly, in Figure \ref{F9}(c), we see that all oscillation rates except for $\beta_i = 0$ have the same survival probability at $t = 1\times10^4$. This suggests that the survival probability is insensitive to the rate of oscillation; rather it is impacted by the current phase of oscillatory \textcolor{black}{evolution} and the magnitude of the oscillation. \textcolor{black}{This dependence on the phase of oscillation has been observed for diffusive particles undergoing oscillatory forcing on fixed domains \cite{menon1992}.} If we consider a fixed minimum domain size and vary the rate of oscillation, as in Figure \ref{F9}(d), we can see this impact clearly. There is an increased sensitivity to the phase of growth if there have been few oscillations, which occurs for $\beta_i$ values with a low magnitude. This sensitivity decreases as the magnitude of the oscillation rate increases. Again, when the oscillating domains align in size (at $t = 1\times10^4$) we see that the rate of oscillation does not affect the survival probability, provided there is any oscillation. In the context of contracting cardiomyocytes, this suggests that the traversal of chemical species across the cell may not be impacted by transient changes to heart rate.

\section{Discussion and conclusions}

Diffusive processes occur on expanding and contracting domains, from the smallest of biological scales \cite{bers2002,zavgorodnya2017} to the largest of cosmological scales \cite{riess1998}. Mathematical investigations have yielded detailed insight into the dynamics of diffusive processes on both single uniformly growing domains and multiple non-growing domains \cite{carr2016,carr2018,carr2019,crampin1999,crampin2002a,landman2003,le2018,klika2017,mclean2004,ross2016,ryabov2015,simpson2015a,simpson2015c,yuste2016}. However, certain diffusive processes of interest, such as drug delivery and cell migration, occur on domains that can grow in a spatially non-uniform manner \cite{binder2008,zavgorodnya2017}. \\

Here we have presented exact solutions to a mathematical model of a diffusive process on multiple growing domains. Each domain exhibits spatially-uniform growth and hence the overall domain can exhibit spatially non-uniform growth. We compare the enumerated exact solution profiles against profiles obtained from repeated realisations of a corresponding lattice-based random walk to ensure that the exact solutions are valid. In all cases, we observe that the derived exact solutions accurately describe the average random walk behaviour. From our exact solutions, we derive expressions for relevant time-varying statistics such as the survival probability and splitting probabilities. We reveal how domain-specific model parameters influence these statistics, including how the interplay between domain growth rates, domain-specific diffusivities and initial location drive long-term survival. Finally, we show how oscillating domain \textcolor{black}{evolution} enhances diffusion. Intriguingly, the rate of oscillation (provided it is non-zero) does not influence the survival probability, provided the comparison is made at a point of time such that the domains have completed an integer number of oscillations. Instead, the magnitude of the oscillation is the key factor influencing the survival probability. This result has interesting implications for biological phenomena that exhibit oscillatory \textcolor{black}{domain evolution}. For example, in the context of cardiomyocytes, this result suggests that the internal diffusion of chemical species should be robust to changes in heart rate, such as those due to exercise or stress. \\

The work presented here may be extended in several directions. We have only considered linearly diffusive processes (i.e. random motion of the agents). However, many processes exhibit characteristics attributable to nonlinear diffusion, such as compact support and a finite rate of spread \cite{johnston2017,li2022,vazquez2007}. These characteristics are particularly relevant for growing domains, as the relationship between the rate of spread and the rate of domain growth dictates whether it is possible for agents to reach the external domain boundary. It is instructive to determine which nonlinear diffusivity functions permit exact solutions following the transformation techniques presented here. Similarly, many biological processes require the inclusion of a reaction term to accurately capture the underlying population dynamics \cite{johnston2017}. It is unclear which classes of reaction terms would admit exact solutions under the transformations examined here; however, it is possible that non-classical transformation techniques may yield analytical progress \cite{bradshaw2019}. \textcolor{black}{Here we have focused on one-dimensional processes. However, for diffusive processes on single growing domains, exact results have been derived for higher dimensions \cite{simpson2015b,johnston2023}. An investigation into the domain geometries that permit exact solutions for multiple growing domains in higher dimensions would be instructive. In the model considered, there are two key timescales on each domain: the timescale of domain growth and the timescale of diffusive motion, which are accounted for via the transformation of the spatial and temporal variables. It would be of interest to investigate whether the approach developed here can be applied in the case where an additional timescale is present due to, for example, chemotactic motion \cite{simpson2006}.}

\section*{Code availability}
\noindent The code used to generate the results presented here can be found on Github at: \\
 https://github.com/DrStuartJohnston/heterogeneous-growing-domains.

\section*{Acknowledgements}
\noindent The authors would like to acknowledge the organisers of the MATRIX workshop ``The mathematics of tissue dynamics,'' where this research project commenced.

\section*{Funding}
\noindent This work was in part supported by the Australian Research Council (STJ: DE200100988, MJS: DP200100177).

\bibliography{Johnston.bib}
\newpage
\section*{Appendix A. Additional solutions}
\begin{figure}[h!]
\begin{center}
\includegraphics[width=0.6\textwidth]{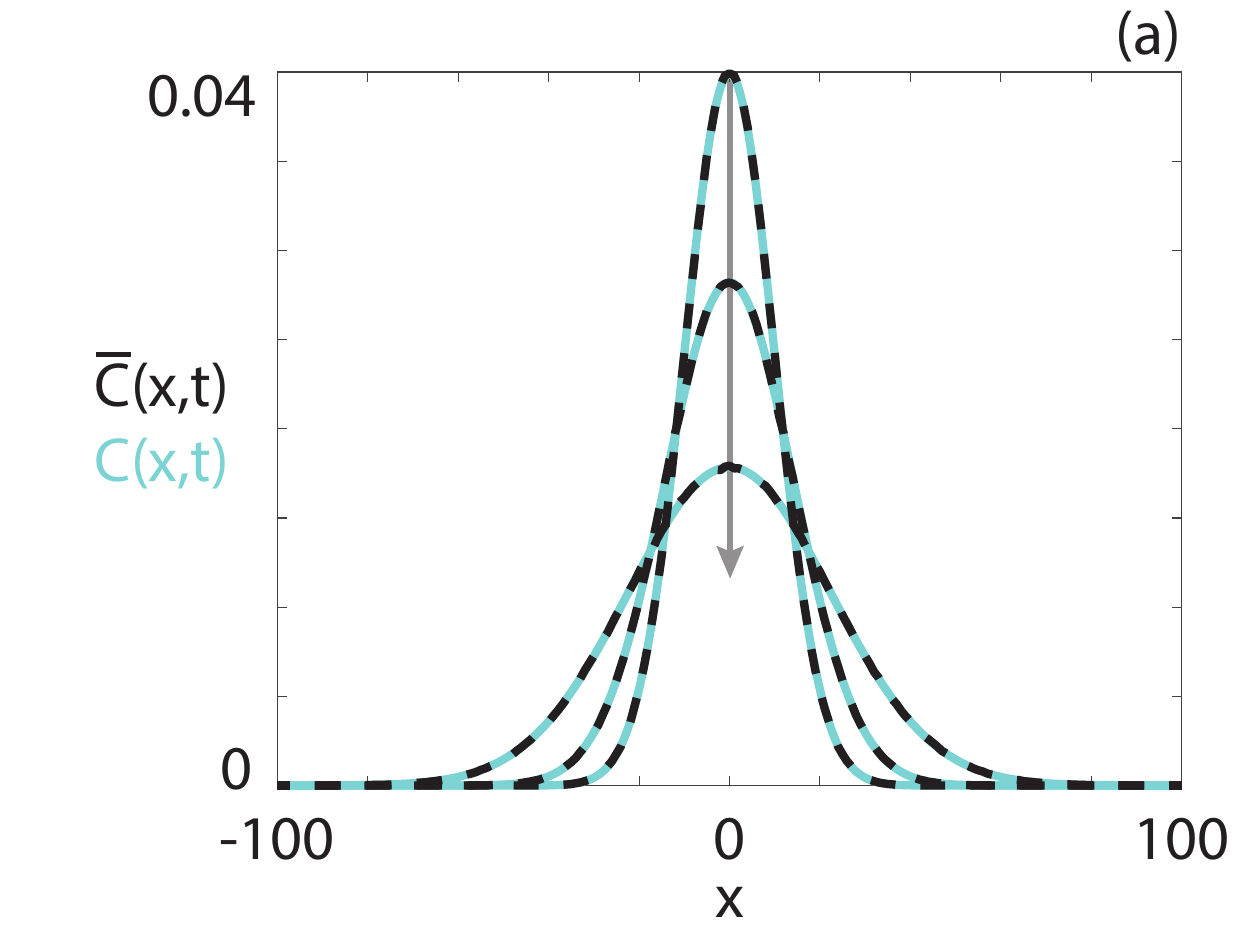}
\caption{Comparison between the average behaviour in the lattice-based random walk $\overline{C}(x,t)$ (black, dashed) and the exact solution $C(x,t)$ as defined in Equation \eqref{Eq:GrowingDomainSolution} (cyan) for a single fixed domain. Parameters used are $D = 0.5$, $L = 100$, $N = 1000$.  Solution profiles are presented at $t = 100$, $t = 250$ and $t = 500$ The arrow indicates the direction of increasing time. Average random walk behaviour is obtained from 5000 identically-prepared realisations of the random walk.}
\label{SI_F1}
\end{center}
\end{figure}

\begin{figure}
\begin{center}
\includegraphics[width=0.6\textwidth]{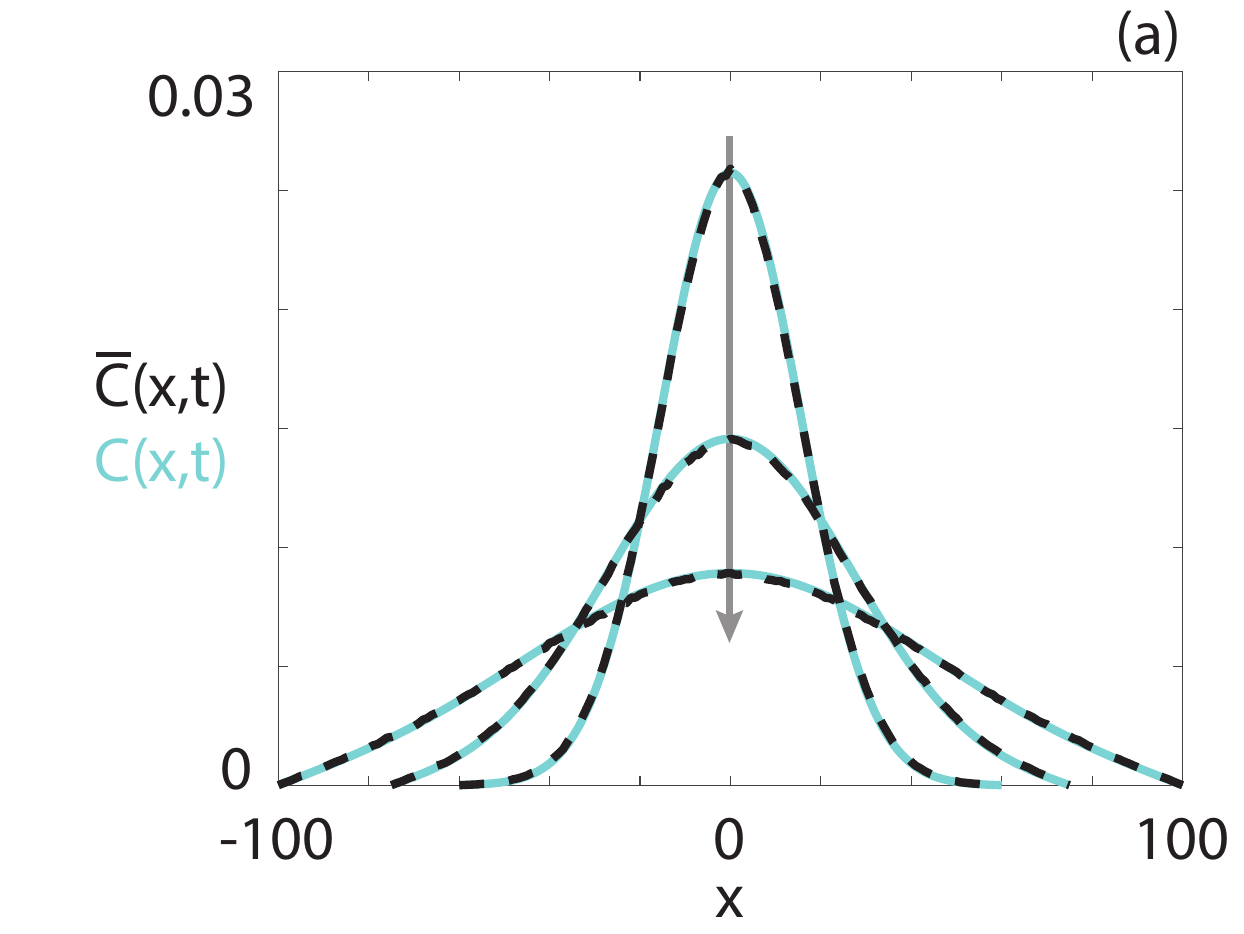}
\caption{Comparison between the average behaviour in the lattice-based random walk $\overline{C}(x,t)$ (black, dashed) and the exact solution $C(x,t)$ as defined in Equation \eqref{Eq:GrowingDomainSolution} (cyan) for a single linearly growing domain (Equations \eqref{Eq:LinearDomainGrowth}-\eqref{Eq:LinearTimeTransformation}). Parameters used are $D = 0.5$, $L = 50$, $N = 1000$, $\beta = 0.05$. Solution profiles are presented at $t = 200$, $t = 500$ and $t = 1000$. The arrow indicates the direction of increasing time. Average random walk behaviour is obtained from 5000 identically-prepared realisations of the random walk.}
\label{SI_F2}
\end{center}
\end{figure}

\begin{figure}
\begin{center}
\includegraphics[width=0.9\textwidth]{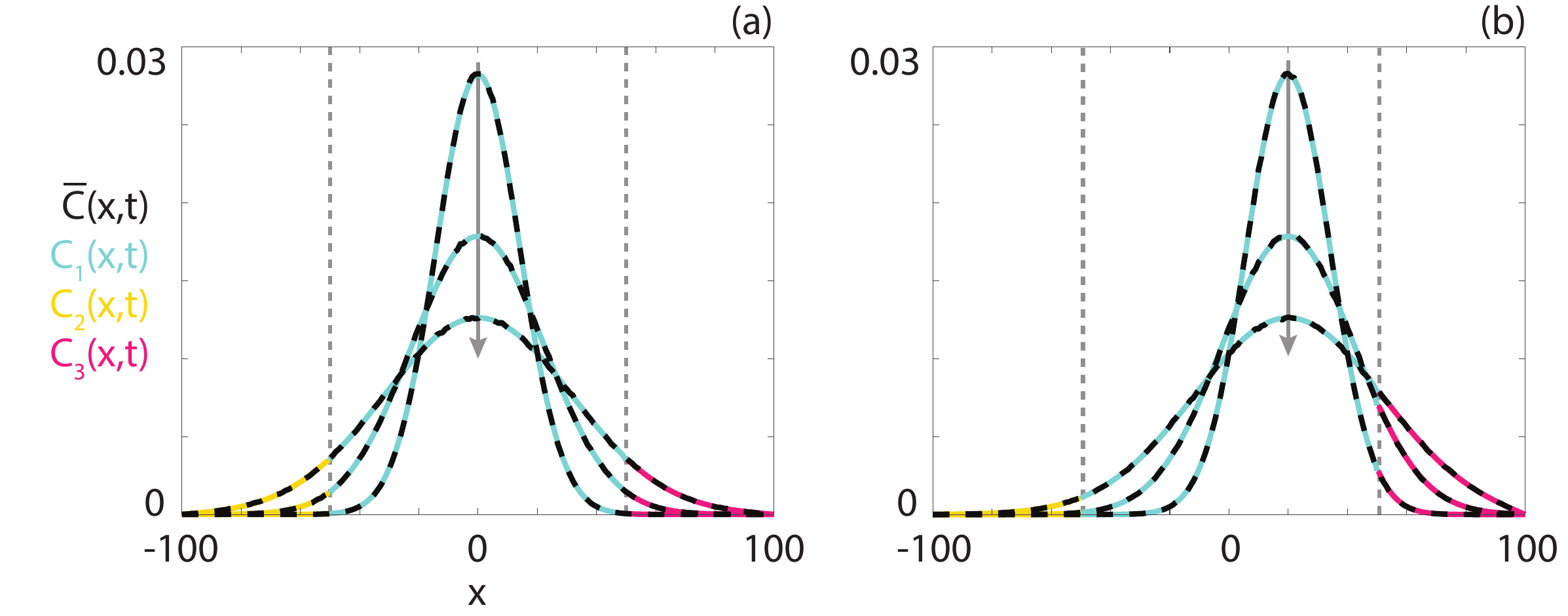}
\caption{Comparison between the average behaviour in the lattice-based random walk $\overline{C}(x,t)$ (black, dashed) and the exact solutions $C_1(x,t)$ (cyan), $C_2(x,t)$ (orange) and $C_3(x,t)$ (pink) as defined in Equations \eqref{Eq:MultipleGrowingDomainsSolution1}-\eqref{Eq:MultipleGrowingDomainsSolution4} for multiple fixed domains with domain-independent diffusivities. Parameters used are $D_1 = D_2 = 0.5$, $L_1 = L_2 = 50$, $N = 1000$, (a)$x_0 = 0$, (b) $x_0 = 20$. Solution profiles are presented at $t = 200$, $t = 500$ and $t = 1000$. The arrow indicates the direction of increasing time. Average random walk behaviour is obtained from 5000 identically-prepared realisations of the random walk.}
\label{SI_F3}
\end{center}
\end{figure}

\begin{figure}
\begin{center}
\includegraphics[width=1.0\textwidth]{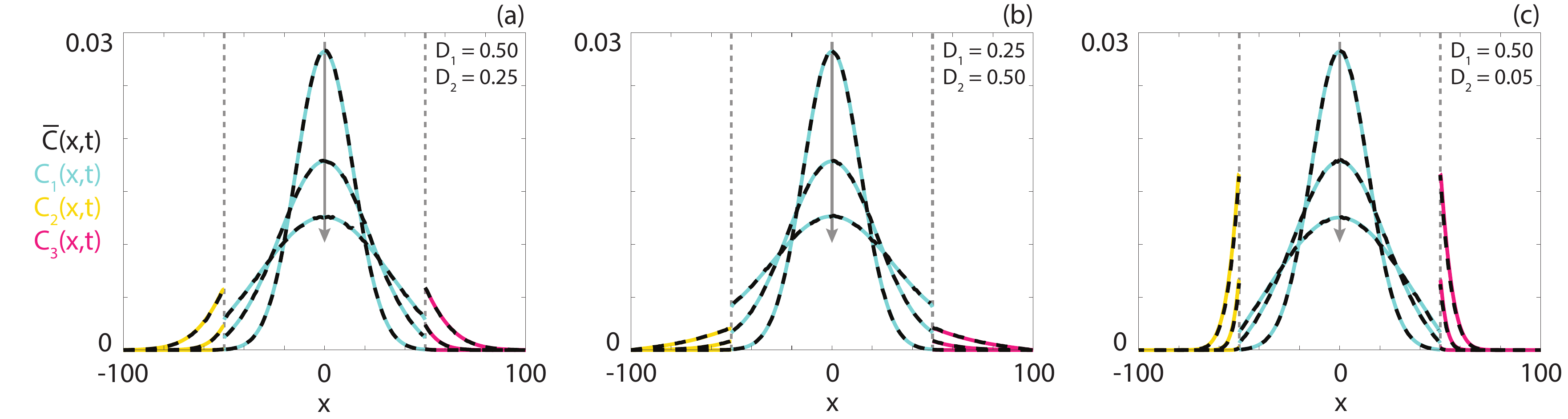}
\caption{Comparison between the average behaviour in the lattice-based random walk $\overline{C}(x,t)$ (black, dashed) and the exact solutions $C_1(x,t)$ (cyan), $C_2(x,t)$ (orange) and $C_3(x,t)$ (pink) as defined in Equations \eqref{Eq:MultipleGrowingDomainsSolution1}-\eqref{Eq:MultipleGrowingDomainsSolution4} for multiple fixed domains and domain-independent diffusivities. Parameters used are $L_1 = L_2 = 50$, $N = 1000$, $x_0 = 0$, (a)  $D_1 = 0.5$, $D_2 = 0.25$, (b)  $D_1 = 0.25$, $D_2 = 0.5$, (c)  $D_1 = 0.5$, $D_2 = 0.05$. Solution profiles are presented at $t = 200$, $t = 500$ and $t = 1000$. The arrow indicates the direction of increasing time. Average random walk behaviour is obtained from 5000 identically-prepared realisations of the random walk.}
\label{SI_F4}
\end{center}
\end{figure}

\begin{figure}
\begin{center}
\includegraphics[width=0.8\textwidth]{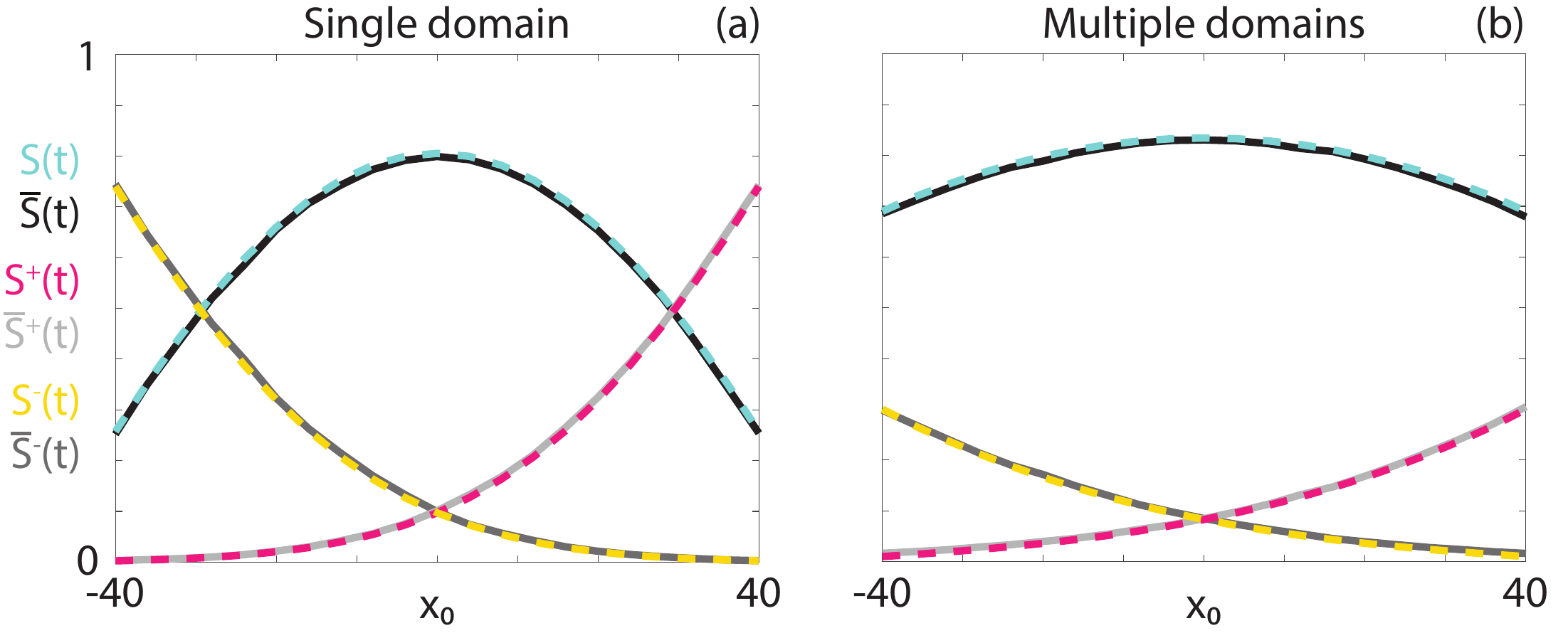}
\caption{Comparison between the survival (black, cyan), left splitting (dark grey, orange) and right splitting (light grey, pink) probabilities obtained from the lattice-based random walk (solid) and the exact solutions defined in (a) Equations \eqref{Eq:LeftSplitSingleDomain}-\eqref{Eq:SurvivalSingleDomain}, (b) Equations \eqref{Eq:RightSplitMultipleDomains}-\eqref{Eq:SurvivalMultipleDomains} (dashed) for (a) one linearly growing domain and (b) two exponentially growing domains (Equations \eqref{Eq:LinearDomainGrowth}-\eqref{Eq:LinearTimeTransformation}). Parameters used are $N = 1000$, (a) $L_1(0) = 50$, $D_1 = 0.5$ and $\beta_1 = 0.05$ and (b) $L_1(0) = L_2(0) = 50$, $D_1 = D_2 = 0.5$ and $\beta_1 = \beta_2 = 0.05$. Probabilities are presented at $t = 10^4$. Average random walk behaviour is obtained from 100 identically-prepared realisations of the random walk.}
\label{SI_F5}
\end{center}
\end{figure}

\begin{figure}
\begin{center}
\includegraphics[width=0.5\textwidth]{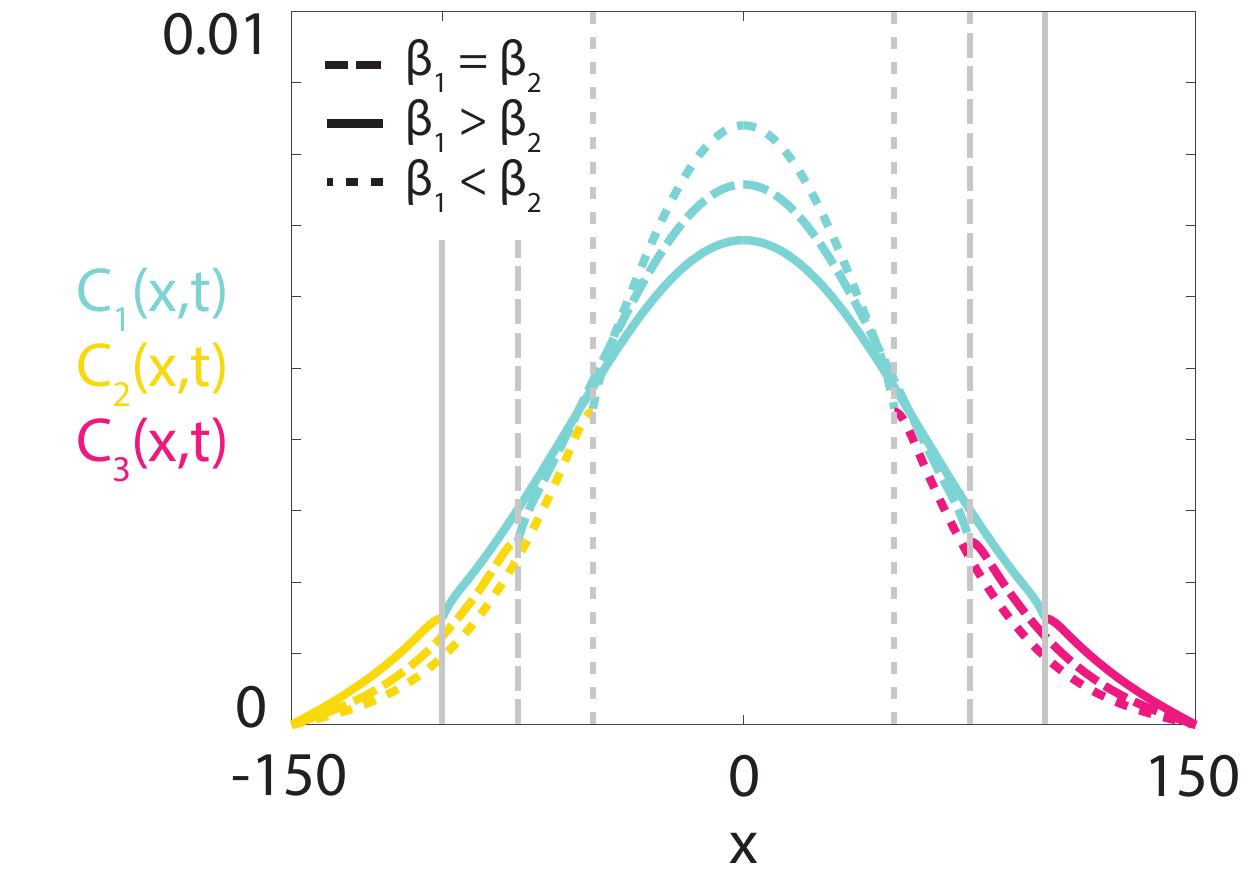}
\caption{\textcolor{black}{Comparison between heterogeneous and homogeneous domain growth. Exact solutions $C_1(x,t)$ (cyan), $C_2(x,t)$ (orange) and $C_3(x,t)$ (pink) as defined in Equations \eqref{Eq:MultipleGrowingDomainsSolution1}-\eqref{Eq:MultipleGrowingDomainsSolution4} for multiple exponentially growing domains (Equations \eqref{Eq:ExpDomainGrowth}-\eqref{Eq:ExpTimeTransformation}) with domain-independent diffusivities. Parameters used are $D_1 = D_2 = 0.5$, $L_1(0) = L_2(0) = 150$, $L_{1,\text{min}} = L_{2,\text{min}} = 25$ $\beta_1 = \beta_2 = -\ln(50/125)/5000$, (dashed lines), $\beta_1 = -\ln(25/125)/5000$, $\beta_2 = -\ln(75/125)/5000$ (solid lines), $\beta_1 = -\ln(75/125)/5000$, $\beta_2 = -\ln(25/125)/5000$ (dashed lines). Solution profiles are presented at $t = 5000$. Grey lines correspond to the position of the boundary.}}
\label{SI_F6}
\end{center}
\end{figure}

\subsection{Random walk algorithm}
\begin{algorithm}[h!]
\caption{Pseudocode algorithm for the implementation of the lattice-based random walk.}
\noindent\scalebox{.8}{%
\begin{minipage}[t]{1.2\linewidth}
\begin{algorithmic}
\While{Time $<$ FinalTime AND NumberOfAgents $> 0$}
	\For{$g = 1,\ldots,$ NumberOfDomains}
		\If{Floor\Big(DomainLength$_g$(Time)/LatticeWidth\Big) $>$ \\ \qquad \qquad \qquad Floor\Big(DomainLength$_g$(PreviousTime)/LatticeWidth\Big)}
			\State{$k \gets \text{rand}_{\text{int}}\Big[\sum_{i=1}^{g-1}$ Floor\Big(DomainLength$_i$(PreviousTime)/LatticeWidth\Big), \\ \qquad \qquad \qquad \qquad $\sum_{i=1}^{g}$ Floor\Big(DomainLength$_i$(PreviousTime)/LatticeWidth\Big)$\Big]$}
			\For{$i = 1,\ldots, $ NumberOfAgents}
				\If{$\text{Position}_i \geq k\times$LatticeWidth}
				\State $\text{Position}_i \gets \text{Position}_i +$ LatticeWidth
				\EndIf
			\EndFor
			\State{$k \gets \text{rand}_{\text{int}}\Big[-\sum_{i=1}^{g}$ Floor\Big(DomainLength$_i$(PreviousTime)/LatticeWidth\Big), \\ \qquad \qquad \qquad \qquad $-\sum_{i=1}^{g-1}$ Floor\Big(DomainLength$_i$(PreviousTime)/LatticeWidth\Big)$\Big]$}
			\For{$i = 1,\ldots,$ NumberOfAgents}
				\If{$\text{Position}_i \leq k\times$LatticeWidth}
				\State $\text{Position}_i \gets \text{Position}_i -$ LatticeWidth
				\EndIf
			\EndFor
			\State TotalDomainLength $\gets$ TotalDomainLength + 2$\times$LatticeWidth
		\EndIf
\If{Floor\Big(DomainLength$_g$(Time)/LatticeWidth\Big) $<$ \\ \qquad \qquad \qquad Floor\Big(DomainLength$_g$(PreviousTime)/LatticeWidth\Big)}
			\State{$k \gets \text{rand}_{\text{int}}\Big[\sum_{i=1}^{g-1}$ Floor\Big(DomainLength$_i$(PreviousTime)/LatticeWidth\Big), \\ \qquad \qquad \qquad \qquad $\sum_{i=1}^{g}$ Floor\Big(DomainLength$_i$(PreviousTime)/LatticeWidth\Big)$\Big]$}
			\For{$i = 1,\ldots, $ NumberOfAgents}
				\If{$\text{Position}_i > k\times$LatticeWidth}
				\State $\text{Position}_i \gets \text{Position}_i -$ LatticeWidth
				\EndIf
			\EndFor
			\State{$k \gets \text{rand}_{\text{int}}\Big[-\sum_{i=1}^{g}$ Floor\Big(DomainLength$_i$(PreviousTime)/LatticeWidth\Big), \\ \qquad \qquad \qquad \qquad $-\sum_{i=1}^{g-1}$ Floor\Big(DomainLength$_i$(PreviousTime)/LatticeWidth\Big)$\Big]$}
			\For{$i = 1,\ldots,$ NumberOfAgents}
				\If{$\text{Position}_i < k\times$LatticeWidth}
				\State $\text{Position}_i \gets \text{Position}_i +$ LatticeWidth
				\EndIf
			\EndFor
			\State TotalDomainLength $\gets$ TotalDomainLength - 2$\times$LatticeWidth
		\EndIf
	\EndFor
	\For{$i = 1\ldots $ NumberOfAgents}
		\State $l \gets \text{rand}_{\text{int}}\Big[1, \text{NumberOfAgents}\Big]$
		\For{$ g= 1, \ldots, $ NumberOfDomains}
			\If{Position$_l \in$ Domain$_g$}
				\State $k \gets \text{rand}_{\text{float}}\big(0,1\big)$
				\If{$k < $ MovementProbability$_g/2$}
				\State $\text{Position}_l \gets \text{Position}_l +$ LatticeWidth
				\ElsIf{$k >$ MovementProbability$/2$  AND $k <$ MovementProbability$_g$}
				\State $\text{Position}_l \gets \text{Position}_l -$ LatticeWidth
				\EndIf
				\EndIf
				\EndFor
		\If{$\text{Position}_l >$ TotalDomainLength  OR $\text{Position}_l < -$TotalDomainLength}
			\State $\text{Position}_l \gets \text{NULL}$
			\State NumberOfAgents $\gets$ NumberOfAgents - 1
		\EndIf
	\EndFor
\State PreviousTime $\gets$ Time
\State Time $\gets$ Time $+$ TimestepLength
\EndWhile
\end{algorithmic}
\end{minipage}}
\end{algorithm}

\end{document}